\documentclass[aps,pre,twocolumn,preprintnumbers,
amsmath,amssymb,floatfix,nofootinbib,reprint]
{revtex4-1}

\usepackage{amsmath}
\usepackage{amsfonts}
\usepackage{amssymb}
\usepackage{graphicx}%
\usepackage{dcolumn}%
\usepackage{bm}%
\usepackage{color}
\usepackage{multirow}
\usepackage{hyperref}
\usepackage[caption=false, font=small]{subfig}
\usepackage[utf8]{inputenc}

\begin{document}

%
%

\title{Nuclear excitation by electron capture in optical-laser-generated plasmas}

%
%

\author{Jonas \surname{Gunst}}
\email{jonas.gunst@mpi-hd.mpg.de}
\affiliation{Max-Planck-Institut f\"ur Kernphysik, Saupfercheckweg 1, D-69117 Heidelberg, Germany}

\author{Yuanbin \surname{Wu}}
\email{yuanbin.wu@mpi-hd.mpg.de}
\affiliation{Max-Planck-Institut f\"ur Kernphysik, Saupfercheckweg 1, D-69117 Heidelberg, Germany}

\author{Christoph H. \surname{Keitel}}
\affiliation{Max-Planck-Institut f\"ur Kernphysik, Saupfercheckweg 1, D-69117 Heidelberg, Germany}

\author{Adriana \surname{P\'alffy}}
\email{palffy@mpi-hd.mpg.de}
\affiliation{Max-Planck-Institut f\"ur Kernphysik, Saupfercheckweg 1, D-69117 Heidelberg, Germany}


\date{\today}

%
%
%
%
%
%
%
\begin{abstract}
The process of nuclear excitation by electron capture in plasma environments generated by the interaction of ultra-strong optical lasers with solid-state samples is investigated theoretically.
With the help of a plasma model we perform a comprehensive study of the optimal parameters for most efficient nuclear excitation and determine the corresponding laser setup requirements. We discern between the low-density plasma regime, modeled by scaling laws, and the high-density regime, for which we perform particle-in-cell calculations. As nuclear transition case study we consider the 4.85 keV nuclear excitation starting from the long-lived $^{93\mathrm{m}}$Mo isomer. Our results show that the optimal plasma and laser parameters are sensitive to the chosen observable and that measurable rates of nuclear excitation and isomer depletion of $^{93\mathrm{m}}$Mo should be already achievable at laser facilities existing today.

\end{abstract}


\maketitle

\section{Introduction}


The invention of the laser more than 50 years ago \cite{Maiman1960} revolutionized atomic physics, leading to the better understanding and control of atomic and molecular dynamics.
Covering several frequency scales, intense coherent light sources available today open unprecedented possibilities for the field of laser-matter interactions \cite{DiPiazzaRMP2012} also  beyond atomic physics. Novel x-ray sources as the X-ray Free Electron Laser (XFEL) open for instance new possibilities to drive low-lying electromagnetic transitions in nuclei \cite{Chumakov2018}. On the other hand, high-power optical lasers with their tremendous efficiency in transferring kinetic energy to charged particles may cause formation of plasma  \cite{bookDieter}, the host of complex interactions between photons, electrons, ions and the atomic nucleus. Nuclear excitation in optical laser-generated plasmas has been the general subject of several studies so far \cite{HarstonPRC1999, Gosselin2004, Gosselin2007, Morel2004-local, Morel2004-local-Hg, Morel2010-nonlocal, Comet2015-NEET, KenL2000.PRL, Cowan2000.PRL, Gibbon2005.Book, Spohr2008.NJP, Mourou2011.S,U-experiment,NIF2018}, while the possibilities to induce nuclear transitions to low-lying excited levels using high-energy lasers have been summarized in Refs.~\cite{Andreev2000, Andreev2001, Granja2007-list-nuclei, Renner2008, Gobet2011}. In addition, nuclear excitation mechanisms in cold high-density plasmas generated by the interaction of  XFEL sources with solid-state targets were investigated in Refs.~\cite{GunstPRL2014, GunstPOP2015}.

Nuclear excitation may occur in plasmas via several mechanisms. Apart from direct or secondary photoexcitation, the coupling to the atomic shell via processes such as nuclear excitation by electron transition (NEET) or electron capture (NEEC) \cite{GoldanskiiPLB1976, PalffyCP2010}  may play an important role. In particular, it was shown that as secondary process in the plasma environment, NEEC may exceed the direct nuclear photoexcitation at the XFEL by approximately six orders of magnitude  \cite{GunstPRL2014, GunstPOP2015}. Since the nuclear coupling to the atomic shell is generally speaking very sensitive to the plasma conditions, the question raises whether one can tailor the latter for maximizing the effect of NEEC. While the tunability of the XFEL-generated plasma properties are limited due to the specific properties of x-ray-atoms interaction, high-power optical lasers are able to generate plasmas over a broad parameter region as far as both temperature and plasma density are concerned. In a very recent letter, NEEC was shown to be the dominant nuclear excitation mechanism for a broad parameter range in optical-laser-generated plasmas, and results on tailored optical laser parameters for maximizing NEEC were presented \cite{Wu2018}.

In this work, we present a systematic theoretical study of NEEC in optical-laser-generated plasmas. Having the concrete scenario of ultra-short optical laser-generated plasmas in mind, we develop a plasma model that can be easily applied to any nuclear parameters. Based on this model we deduce the optimal NEEC parameters in terms of plasma temperature and density. Due to the  complexity of the processes involved, we show that the plasma parameters for maximal NEEC rate are  not identical to the ones that determine the maximal number of excited nuclei. We then further investigate how the optimal NEEC parameters in the temperature-density landscape can be accessed by a short and intense laser pulse considering  various experimental conditions such as laser intensity, wavelength, pulse duration or pulse energy. We discern in our treatment between two cases, the low-density plasma regime, modeled by the scaling law, and the high-density regime, where we use dedicated particle-in-cell (PIC) simulations.

As case study we consider the 4.85 keV nuclear transition starting from a long-lived excited state of  $^{93}$Mo at 2.4 MeV. Such states are also known as nuclear isomers \cite{WalkerReview2016} and have been the subject of increased attention due to the potential storage of large amounts of energy over long periods of time \cite{WalkerN1999, AprahamianNP2005, BelicPRL1999, CollinsPRL1999, BelicPRC2002, CarrollLPL2004, PalffyPRL2007}.
For  $^{93\mathrm{m}}$Mo, the additional 4.85 keV nuclear excitation leads to the depletion of the isomer and release on demand of the stored energy. Apart from this appealing scenario, $^{93\mathrm{m}}$Mo is interesting also because of  the recently reported observation of isomer depletion via NEEC of the 4.85 keV transition \cite{Chiara2018}. Our results show that for high electron densities, NEEC is actually the dominant nuclear excitation channel for $^{93\mathrm{m}}$Mo, surpassing by orders of magnitude photoexcitation in the plasma. Surprisingly, a six orders of magnitude increase in the number of excited nuclei can be achieved employing a high-power optical laser compared to the previously investigated case of XFEL-generated plasmas. The calculated
maximal number of depleted isomers for realistic laser
setup parameters appears to have reached values large enough to be observable in an experiment.  Although still far from the final goal, this is a further
milestone on the way to the realization of controlled energy
storage and release via nuclear isomers.

This paper is structured as follows. After introducing the theory for NEEC and nuclear photoexcitation in a plasma environment and the employed plasma model in Section \ref{sec:theory}, we start by investigating the optimal plasma conditions for NEEC in terms of electron density and temperature based on a simplified model for spherical plasmas in Section \ref{sec:spherical}. As part of this Section, we discuss the influence of ionization potential depression, the expected contribution from photoexcitation and a hydrodynamic model for plasma expansion. Section \ref{sec:laser} is then devoted to optical laser-generated plasmas. Since the plasma generation involves different processes depending on the electron density, we divide this Section into two parts, \ref{sec:low} for low density and \ref{sec:high} for high density plasmas, respectively. As a main result we evaluate the optimal laser parameters for low-density scenarios at the example of high-power laser facilities. Surprisingly, our analysis of a PIC simulation for high electron densities shows that similar nuclear excitation numbers can be achieved with 100 J lasers available at many facilities around the world. The paper ends with concluding remarks.

\section{Theoretical approach}
\label{sec:theory}

This Section introduces the theoretical approach used for describing nuclear excitation in the plasma.
After first general considerations on the setup, we will sketch our calculations  for NEEC and photoexcitation rates in the plasma and outline the used plasma model. Atomic units $\hbar=m_{\text{e}}=e=4\pi\varepsilon_0=1$ are used throughout subsections \ref{sec:theory_neec} and \ref{sec:photoexc}.

\subsection{General considerations}

We investigate the interaction of ultra-strong lasers with a solid-state target. The strong electromagnetic field of optical lasers leads to field ionization and acceleration of the electrons in the target. The accelerated hot electrons can lead to further ionization of the target via collisional ionization.
Since the laser wavelength is in the optical range, the electronic heating occurs essentially at the surface of the material. The hot electrons produced by the laser are accelerated away from the interaction region generating an electric field due to the charge separation. 
Attracted by this electric field the ions subsequently follow the hot electrons, resulting in the formation of a neutral plasma surrounding the target. The time scale of the plasma generation process is on the order of the pulse duration (plus the time for the acceleration of the ions). Moreover, near the focal spot of the laser the plasma can be considered as uniform in terms of electron density and temperature.

Looking in more detail at the start of the laser-target interaction, there will be a pre-pulse leading to the generation of a pre-plasma. The main pulse thus interacts with this cold pre-plasma instead of the initial target. Consequently, the final plasma consists of a cold (coming from the pre-pulse) and a hot electron distribution (coming from the main pulse). The cold electron distribution stays essentially in the region around the interaction point with the target. As far as we consider thin targets ($\sim \mu$m) this volume is small in comparison to the total plasma volume, such that the cold electrons can be neglected.

In order to reach high electron densities ($\sim 10^{22}$ cm$^{-3}$) thicker targets need to be considered. Analogously to the case of thin targets, the electromagnetic laser field accelerates the electrons of the target surface into the inner region of the target. These hot electrons then subsequently lead to further collisional ionization events inside the target which can result in a very dense plasma in this region \cite{SentokuPOP2007, HuangPOP2013}. However, the absorption fraction $f$ seems to be effectively much lower for this heating mechanism in comparison to the thin target case.

For our particular study of nuclear excitation in plasmas, we consider a strong optical laser that interacts with a solid-state target containing a  fraction of nuclei in the isomeric state.  NEEC and/or photoexcitation  may occur in the generated plasma.
In the resonant process of NEEC, a free electron recombines into a vacant bound atomic state with the simultaneous excitation of the nucleus. The isomers can then be excited to a trigger state which rapidely decays to the nuclear ground state and releases the energy ``stored'' in the isomer. For $^{93\mathrm{m}}$Mo the stored energy corresponds to the isomeric state at  2424 keV.  A further 4.85 keV excitation leads to the fast release within approx. 4 ns via a decay cascade containing a 1478 keV photon which could serve as signature of the isomer depletion. Recent experimental results on the isomer depletion of $^{93\mathrm{m}}$Mo in a channeling  setup have deduced a rather high NEEC rate of the  4.85 keV transition \cite{Chiara2018} encouraging further studies on this nuclear transition.

\subsection{NEEC in plasma environments}
\label{sec:theory_neec}

In the plasma, free electrons with different kinetic energies are available. At the NEEC resonant energy, electrons may recombine into ions leading to nuclear excitation.  
 The resonance bandwidth is determined by a  narrow Lorentz profile.
Since the kinetic energy of free electrons in a plasma is distributed over a wide range, many resonant NEEC channels may exist. In the following we will shortly describe how such a situation can be handled theoretically in terms of reaction rates.

In order to restrict the number of possible initial electron configurations, for a lower-limit estimate,
we consider in the following only NEEC into ions which are in their electronic ground states. In this case, the initial electronic configuration $\alpha_{\text 0}$ is uniquely identified by the charge state number $q$ before electron capture.
In the isolated resonance approximation, the total NEEC reaction rate in the plasma can be written as a summation over all charge states $q$ and all capture channels $\alpha_{\text{d}}$,
\begin{equation}
\lambda_{\text{neec}}(T_{\text{e}}, n_{\text{e}}) = \sum_q \sum_{\alpha_{\text{d}}} P_q(T_{\text{e}}, n_{\text{e}}) \lambda^{q,\alpha_{\text{d}}}_{\text{neec}}(T_{\text{e}}, n_{\text{e}})\ .
\label{eq:neec-plasma.total}
\end{equation}
Here, $P_q$ is the probability to find in the plasma ions in the charge state $q$ as a function of electron temperature $T_{\text{e}}$ and density $n_{\text{e}}$.
The partial NEEC rate into the capture level $\alpha_{\text{d}}$ of an ion in the charge state $q$ can be expressed by the convolution over the electron energy $E$ of the  single-resonance NEEC cross section $\sigma^{\text{i} \rightarrow \text{d}}_{\text{neec}}$ and the free-electron flux $\phi_{\text{e}}$,
\begin{equation}
\lambda^{q,\alpha_{\text{d}}}_{\text{neec}}(T_{\text{e}}, n_{\text{e}}) = \int dE \, \sigma_{\text{neec}}^{\text{i} \to \text{d}}(E) \phi_{\text{e}}(E, T_{\text{e}}, n_{\text{e}})\ .
\label{eq:neec-plasma.partial}
\end{equation}
The NEEC cross section $\sigma^{\text{i} \rightarrow \text{d}}_{\text{neec}}$ as a function of the free electron energy $E$ is proportional to a Lorentz profile $L_{\text{d}}(E-E_{\text{d}})$ centered on the resonance energy $E_{\text{d}}$. The width of the resonance is typically determined by the nuclear linewidth  of approx. 100 neV. Since over this energy scale  $\phi_{\text{e}}$ can safely be considered as constant, we can approximate the Lorentz profile by a Dirac-delta-like function.
The match between the resonance energy $E_{\text{d}}$ and the functional temperature dependence of the electron flux $\phi_{\text{e}}(T_{\text{e}})$ determines the quantitative contribution of the individual NEEC channels.

The electron flux $\phi_{\text{e}}$ in the plasma can be written as the product of the density of states $g_{\text{e}}(E)$, the Fermi-Dirac distribution $f_{\text{FD}}(E, T_{\text{e}}, n_{\text{e}})$ for a certain electron temperature $T_{\text{e}}$ and the velocity $v(E)$,
\begin{equation}
\phi_{\text{e}}(E, T_{\text{e}}, n_{\text{e}})\, dE = g_{\text{e}}(E) f_{\text{FD}}(E, T_{\text{e}}, n_{\text{e}}) v(E)\, dE\ .
\label{eq:electron.flux.definition}
\end{equation}
The temperature dependence of $\phi_{\text{e}}$ is only included in the Fermi-Dirac statistics $f_{\text{FD}}$ \cite{bookOxenius}. The density of states as well as the velocity are determined by considering the relativistic dispersion relation for the free electrons.
The electronic chemical potential $\mu_{\text{e}}$  occurring in the expression of $f_{\text{FD}}$ is fixed by adopting the normalization
\begin{equation}
    \int dE \, g_{\text{e}}(E) f_{\text{FD}}(E, T_{\text{e}}, n_{\text{e}}) = n_{\text{e}}\ .
    \label{eq:electron.flux.norm}
\end{equation}
Thus, the electron flux in the plasma depends on both the electron temperature $T_{\text{e}}$ and the density $n_{\text{e}}$.

The theoretical formalism for the calculation of the NEEC cross section $\sigma_{\text{neec}}$ has been presented elsewhere ~\cite{PalffyPRA2006, PalffyPRA2007, GunstPRL2014, GunstPOP2015}. The cross section is connected to the microscopic NEEC reaction rate $Y_{\text{neec}}$ via
\begin{equation}
    \sigma_{\text{neec}}^{\text{i} \rightarrow \text{d}}(E) = \frac{2 \pi^2}{p^2} Y_{\text{neec}}^{\text{i} \rightarrow \text{d}} L_{\text{d}}(E-E_{\text{d}}) \ ,
    \label{eq:Y_neec}
\end{equation}
with $p$ the free electron momentum. Substituting Eq.~\eqref{eq:Y_neec} into Eq.~\eqref{eq:neec-plasma.partial}, the integral over the kinetic electron energy $E$ can be solved by assuming that the free electron momentum $p$ and the NEEC rate $Y_{\text{neec}}$ are constant over the width of the Lorentz profile $L_{\text{d}}(E-E_{\text{d}})$, which is the case for a wide spectrum of isotopes.  Eq.~\eqref{eq:neec-plasma.partial} then simplifies to
\begin{equation}
\lambda^{q,\alpha_{\text{d}}}_{\text{neec}}(T_{\text{e}}, n_{\text{e}}) = \frac{2 \pi^2}{p^2} Y_{\text{neec}}^{\text{i} \rightarrow \text{d}} \phi_{\text{e}}(E_{\text{d}}, T_{\text{e}}, n_{\text{e}})\ .
\label{eq:neec-plasma.partial2}
\end{equation}

The total NEEC rate $\lambda_{\text{neec}}$ in Eq.~(\ref{eq:neec-plasma.total}) is therefore strongly dependent on the available charge states and free electron energies which both are dictated by the plasma conditions.
Taking the spatial and temporal plasma evolution into account, the total NEEC excitation number $N_{\text{exc}}$ is connected to the rate $\lambda_{\text{neec}}$ via
\begin{equation} \label{eq:N_exc}
    N_{\text{exc}} = \int_{V_{\text{p}}}\! d^3\bold{r} \int\! dt\, n_{\text{iso}}(\bold{r},t) \, \lambda_{\text{neec}}(T_{\text{e}}, n_{\text{e}};\bold{r},t)\ ,
\end{equation}
where $n_{\text{iso}}$ denotes the number density of isomers present in the plasma.
For the further quantitative estimate of the occurred nuclear excitation,
relevant factors are the interaction time and the volume over which the interaction takes place, considered to be the plasma volume $V_{\text{p}}$. These aspects are detailed in subsection \ref{sec:plasma}.

\subsection{Resonant nuclear photoexcitation in plasma}
\label{sec:photoexc}

Instead of undergoing NEEC, the nucleus can also be excited by the absorption of a photon which has to be on resonance with the nuclear transition energy $E_{\text{n}}$.
Analog to Eqs.~\eqref{eq:neec-plasma.total} and \eqref{eq:neec-plasma.partial}, the excitation rate via photons in the plasma can be expressed as
\begin{equation}
\lambda_{\gamma}(T_{\text{e}}, n_{\text{e}}) = \int \sigma_{\gamma}^{\text{i} \rightarrow \text{d}}(E) \phi_{\gamma}(E,T_{\text{e}},n_{\text{e}})\, dE \ ,
\label{eq:photoexc-rate}
\end{equation}
with the nuclear photoexcitation cross section $\sigma_{\gamma}^{\text{i} \rightarrow \text{d}}(E) = \frac{2 \pi^2}{k^2} A_{\gamma}^{\text{i} \rightarrow \text{d}} L_{\text{d}}(E-E_{\text{n}})$,
where $A_{\gamma}^{\text{i} \rightarrow \text{d}}$ represents the corresponding rate.
For the calculation of the photoexcitation rate, we have adopted the formalism from Ref.~\cite{RingSchuck} that connects $A_{\gamma}$ with so-called reduced nuclear transition probabilities. For the latter we employ experimental data and/or nuclear model calculations later on.

In general, the photon flux and hence the photoexcitation rate in the plasma depend on the prevailing plasma conditions represented by electron temperature and density in Eq.~\eqref{eq:photoexc-rate}. In order to evaluate this dependence further, we employ two models for the photon flux $\phi_{\gamma}$ in the following.

First, we assume the photons to be in thermodynamic equilibrium (TDE) with the electrons, such that a blackbody distribution is applicable, resulting in the density-independent photon flux
\begin{equation}
    \phi_{\gamma}^{\text{TDE}}(E,T_{\text{e}})\, dE = c\, g_{\gamma}(E) f_{\text{BE}}(E, T_{\text{e}})\, dE .
    \label{eq:ph-flux}
\end{equation}
Here, $c$ is the speed of light,  $g_{\gamma}$ represents the photonic density of states and $f_{\text{BE}}$ denotes the Bose-Einstein distribution \cite{bookOxenius}, respectively. Substituting Eq.~\eqref{eq:ph-flux} into Eq.~\eqref{eq:photoexc-rate} leads to the photoexcitation rate under TDE conditions
\begin{equation}
\lambda_{\gamma}^{\text{TDE}}(T_{\text{e}})
= \frac{2 \pi^2}{k^2} A_{\gamma}^{\text{i} \rightarrow \text{d}} \phi_{\gamma}^{\text{TDE}}(E_{\text{n}},T_{\text{e}})\ .
\label{eq:photoexc-rate-blackbody}
\end{equation}
In the derivation of Eq.~\eqref{eq:photoexc-rate-blackbody}, the Lorentzian profile has been approximated by a Dirac-delta-like resonance since the nuclear transition width is for the considered plasma temperatures much smaller than the energy region over which the photon flux varies significantly.

As a second model, we considered the process of bremsstrahlung as a potential photon source in the plasma. According to Ref.~\cite{HarstonPRC1999}, the photon flux emitted via bremsstrahlung evaluates to
\begin{equation}
    \begin{split}
        & \phi_{\gamma}^{\text{B}}(E, E_{\text{e}}, T_{\text{e}}, n_{\text{e}})\, dE\, dE_{\text{e}} \\
        & \quad\quad = t_{\text{i}} \,\left(\frac{d\sigma_{\text{B}}(E_{\text{e}})}{dE}\right)\, \phi_{\text{e}}(E_{\text{e}},     T_{\text{e}}, n_{\text{e}})\, dE\, dE_{\text{e}},
    \end{split}
    \label{eq:ph-flux-brems}
\end{equation}
where $d\sigma_{\text{B}}(E_{\text{e}}) / dE$ denotes the bremsstrahlung cross section differential in the emitted photon energy $E$ and $t_{\text{i}}$ represents the target thickness given in atoms per area. For the calculations later on we consider $t_{\text{i}} = n_{\text{i}} R_{\text{p}}$, where $n_{\text{i}}$ represents the ion number density in the plasma and $R_{\text{p}}$ the plasma radius.
Employing Eq.~\eqref{eq:ph-flux-brems}, the photoexcitation rate in the plasma with photons emitted via bremsstrahlung is given by
\begin{equation}
    \lambda_{\gamma}^{\text{B}}(T_{\text{e}}, n_{\text{e}}) =  \frac{2 \pi^2}{k^2} A_{\gamma}^{\text{i} \rightarrow \text{d}} \int \phi_{\gamma}^{\text{B}}(E_{\text{n}}, E_{\text{e}}, T_{\text{e}}, n_{\text{e}})\, dE_{\text{e}}\ ,
\label{eq:photoexc-rate-brems}
\end{equation}
where the same approximation as for the blackbody spectrum has been used to solve the integration over the kinetic electron energy $E$. The photon flux on resonance occurring in Eq.~\eqref{eq:photoexc-rate-brems} is determined by
\begin{equation}
    \phi_{\gamma}^{\text{B}}(E_{\text{n}}, E_{\text{e}}, T_{\text{e}}, n_{\text{e}})
    =
    \left(\frac{d\sigma_{\text{B}}(E_{\text{e}})}{dE}\right)_{E=E_{\text{n}}} \phi_{\text{e}}(E_{\text{e}}, T_{\text{e}}, n_{\text{e}})\ .
\end{equation}

Replacing $\lambda_{\text{neec}}$ by the corresponding photoexcitation rate [Eq.~\eqref{eq:photoexc-rate-blackbody} or \eqref{eq:photoexc-rate-brems}] in Eq.~\eqref{eq:N_exc}, the total number of excited nuclei via resonant nuclear photoexcitation can be evaluated. A comparison of the NEEC and resonant photon absorption rates in the plasma is presented in
Section \ref{sec:spherical_photo} for a variety of plasma conditions.

\subsection{Plasma model}
\label{sec:plasma}

For the plasma modeling part here and in the following,  SI units with $k_{\rm{B}}=1$ are adopted, unless for some quantities the units are explicitely given.


\subsubsection{General model for spherical plasmas}
\label{sec:plasma-general}

In order to get a general idea of the number of excited nuclei in the plasma, we first disregard the exact  target  heating processes and assume in a first approximation a spherical plasma with homogeneous electron temperature $T_{\text{e}}$ and density $n_{\text{e}}$ over the plasma lifetime $\tau_{\text{p}}$. With that the total number of excited nuclei determined in Eq.~\eqref{eq:N_exc} evaluates to
\begin{equation} \label{eq:N_exc_simple}
    N_{\text{exc}} = N_{\text{iso}} \lambda_{\text{neec}}(T_{\text{e}}, n_{\text{e}}) \tau_{\text{p}}\ ,
\end{equation}
with $N_{\text{iso}}$ being the number of isomers in the plasma. $N_{\text{iso}}$ can be estimated introducing the isomer fraction embedded in the original solid-state target $f_{\text{iso}}$,
\begin{equation}
    N_{\text{iso}} = f_{\text{iso}} n_{\text{i}} V_{\text{p}}\ ,
\end{equation}
where $n_{\text{i}}$ stands for the ion number density in the plasma and the plasma volume $V_{\text{p}}$ is given by $V_{\text{p}} = \frac{4}{3} \pi R_{\text{p}}^3$ with the plasma radius $R_{\text{p}}$. In neutral plasmas, the ion and electron densities are related via the average charge state $\bar{Z}$,
\begin{equation}
    n_{\text{i}} = n_{\text{e}} / \bar{Z}\ .
\end{equation}

In the case of ${}^{93\text{m}}$Mo isomer triggering, an isomer fraction of $f_{\text{iso}} \approx  10^{-5}$ embedded in solid-state Niobium foils can be generated by intense ($\ge 10^{14}$ protons/s) beams  \cite{GunstPRL2014} via the $^{93}_{41}$Nb(p,n)$^{93\mathrm{m}}_{\phantom{m} 42}$Mo reaction \cite{exfor}.

Moreover, the plasma lifetime $\tau_{\text{p}}$ occurring in Eq.~\eqref{eq:N_exc_simple} can be approximated for spherical plasmas by following an estimate for spherical clusters \cite{KrainovS2002}.
The time scale after which the plasma's spatial dimension is approximatively doubled is given as a function of plasma radius, electron temperature and average charge state by
\begin{equation} \label{eq:tau}
  \tau_{\text{p}} = R_{\text{p}} \sqrt{m_{\text{i}}/(T_{\text{e}} \bar{Z})}\ ,
\end{equation}
with the ion mass $m_{\text{i}}$. Note that $\tau_{\text{p}}$ is implicitly also influenced by the electron density $n_{\text{e}}$ due to the dependence $\bar{Z}(T_{\text{e}},n_{\text{e}})$.

Based on the expression of the plasma lifetime $\tau_{\text{p}}$ in Eq.~\eqref{eq:tau}, the total number of excited nuclei in the plasma can be estimated. This approximative approach is easily applicable to other nuclear transitions and provides many instructive insights to plasma-mediated nuclear excitations as shown later on considering the example of ${}^{93\text{m}}$Mo triggering.
In order to test the validity of the plasma lifetime approach, we perform a comparison with results from a hydrodynamic model for the plasma expansion.

\subsubsection{Hydrodynamic expansion}
\label{sec:hydro}

Following the analysis in Ref.~\cite{GunstPOP2015}, we consider a more detailed hydrodynamic model for the plasma expansion by a quasi-neutral expansion of spherical clusters as studied in the context of the intense optical laser pulses interaction with spherical clusters \cite{KrainovS2002, DitmirePRA1996}. During the expansion, the plasma is assumed to maintain a uniform (but decreasing) density throughout the plasma sphere while the electron temperature decreases with the adiabatic expansion of the plasma,
\begin{equation}
  \frac{3}{2} n_{\text{e,t}} V dT_{\text{e,t}} = -P_{\text{e}} dV,
\end{equation}
where $n_{\text{e,t}}$ is the number density of free electrons, $V = 4 \pi R_{\text{t}}^3/3$ is the volume of the plasma with the radius $R_{\text{t}}$, and $P_{\text{e}} = n_{\text{e,t}} T_{\text{e,t}}$ is the pressure of free electrons. The time-dependent electron temperature and the plasma radius satisfy the relation
\begin{equation}
 T_{\text{e,t}} = T_{\text{e}} \left( \frac{R_{\text{p}}}{R_{\text{t}}} \right)^2,
\end{equation}
where $T_{\text{e}}$ is the initial electron temperature and $R_{\text{p}}$ the initial plasma radius. During the plasma expansion, the electrons lose their thermal energy to the ions resulting into the electron and ion kinetic energies
\begin{eqnarray}
  n_{\text{i,t}} \frac{d T_{\text{i,t}}}{dt} = -n_{\text{e,t}} \frac{d T_{\text{e,t}}}{dt}, \\
  \frac{1}{2} m_{\text{i}} \left( \frac{d R_{\text{t}}}{dt} \right)^2 = \frac{3}{2} T_{\text{i,t}}, \label{eq:tionv}
\end{eqnarray}
where $n_{\text{i,t}}$ is the ion number density, $T_{\text{i,t}}$ is the ion temperature, and $m_{\text{i}}$ is the ion mass. The electron and ion collisions take place on a much shorter time scale than the plasma expansion time, such that we can consider the temperature to be uniform throughout the sphere \cite{KrainovS2002}. 
The equation of plasma expansion is given by
\begin{equation} \label{eq:expaneq}
  m_{\text{i}} \frac{d^2 R_{\text{t}}}{dt^2} = 3 Z \frac{T_{\text{e}} R_{\text{p}}^2}{R_{\text{t}}^3},
\end{equation}
where $Z $ is the ratio of the electron density to the ion density, i.e., the average charge state of the ions in the quasi-neutral limit. Solving Eq.~(\ref{eq:expaneq}) for a fixed $Z = \bar{Z}$ under the condition that the initial speed for cluster expansion is zero,  one obtains the plasma lifetime expression $\tau_{\text{p}}$ in Eq.~(\ref{eq:tau}) as the expansion time for increasing  the plasma radius by a factor 2.

Rate estimates based on the FLYCHK code \cite{FLYCHK-IAEA, FLYCHK2005} show that the time scale to reach the steady state varies from the order of $10$ fs for solid-state density to the order of $10$ ps for low density $~\sim 10^{19}$ cm$^{-3}$ at temperature $\sim 1$ keV (for the plasma constituents under consideration in the present work). For solid-state density plasmas, 10 fs is much shorter than the time interval that we are analyzing for NEEC in this work. For the low density case, 10 ps is normally longer than the laser pulse duration. However, the relevant time scale to compare with is not the pulse duration, but rather the lifetime of the plasma, over which NEEC takes place. As discussed below, such low-density plasmas generated by the laser-target interaction under consideration last a few hundreds ps. Therefore, we can conclude that the time scale for reaching the steady state (with regard to the atomic processes) is much smaller than the expansion time scale  for the plasmas conditions under consideration here. We thus may assume that the steady state with regard to the atomic processes establishes at each time instant during the expansion.

\subsubsection{Laser-induced plasma: scaling law}
\label{sec:plasma-scaling}

Considering the case of a low density (underdense) plasma, which can be generated via the interaction of a strong optical laser with a thin target, the plasma generation process typically evolves in two steps \cite{FuchsNP2006}: (i) a preplasma is formed by the prepulse of the laser; (ii) this preplasma is subsequently heated by the main laser pulse potentially up to keV electron energies.
We model the plasma following the approach in Refs.~\cite{FuchsNP2006, WuAPJ2017}. With the help of a so-called scaling law, the plasma conditions can be mapped to laser parameters, like laser intensity $I_{\text{laser}}$, wavelength $\lambda_{\text{laser}}$, pulse duration $\tau_{\text{laser}}$ and pulse energy $E_{\text{pulse}}$. Assuming a flat-top beam profile and for a fixed focal radius $R_{\text{focal}}$ the laser intensity reads
\begin{equation} \label{eq:I_laser}
    I_{\text{laser}} = \frac{E_{\text{laser}}}{\tau_{\text{laser}} \pi R_{\text{focal}}^2}\ .
\end{equation}
We adopt here at first the widely used ponderomotive scaling law in the non-relativistic limit (sharp-edged profiles),
\begin{equation} \label{eq:SL1}
    T_{\text{e}} \approx 3.6 I_{16} \lambda_{\mu}^2\, ~\text{keV}\ ,
\end{equation}
where $ I_{16}$ is the laser intensity in units of $10^{16}$ W/cm$^2$ and $\lambda_{\mu}$ the wavelength in microns \cite{Brunel1987, BonnaudLPB1991, GibbonPPCF1996}.

Depending on the target  and laser-target interaction conditions, different electron temperature scalings are used in the literature \cite{GibbonPPCF1996}. For comparison we adopt here also a second scaling law  (known as short-scale length profile) \cite{GibbonPPCF1996, GibbonPRL1992}:
\begin{equation} \label{eq:SL2}
    T_{\text{e}} \approx 8 \left( I_{16} \lambda_{\mu}^2 \right)^{1/3}\, ~\text{keV}\ .
\end{equation}

The electron density can be estimated as $n_{\text{e}} = N_{\text{e}}/V_{\text{p}}$ where $N_{\text{e}}$ is the total number of electrons which can be related to the absorbed laser energy $f E_{\text{pulse}}$ via
\begin{equation} \label{eq:absorption}
  N_{\text{e}} = \frac{f E_{\text{pulse}}}{T_{\text{e}}}\ .
\end{equation}
The absorption coefficient $f$ saturates at around 10-15\% for high irradiances and steep density profiles \cite{GibbonPPCF1996}. However, in the case of moderate intensities and intermediate scale lengths (e.g. $I \lambda^2$ =  $10^{16}$ W cm$^{-2}$ $\mu$m$^2$, $L/\lambda \sim 0.1$), the absorption can be much higher with $f$ taking values up to 70\% \cite{GibbonPPCF1996}.

The plasma volume in the case of a laser-generated plasma is given by
\begin{equation} \label{eq:V_p}
    V_{\text{p}} = \pi R_{\text{focal}}^2 d_{\text{p}}\ ,
\end{equation}
where the plasma thickness $d_{\text{p}} = c \tau_{\text{pulse}}$ is determined by  the laser pulse duration $\tau_{\text{pulse}}$.
In contrast to the pure spherical plasma, we consider here a cylindrical geometry with the transversal length dimension determined by the focal radius of the laser $R_{\text{focal}}$ and the longitudinal length scale via the plasma thickness $d_{\text{p}}$.

Nevertheless, for the case of focal radius,  plasma thickness and plasma radius of similar scale, we may again consider the plasma lifetime given by Eq.~\eqref{eq:tau} derived for the spherical plasma model.
For a lower-limit estimate of the nuclear excitation, we use the smallest length scale out of $R_{\text{focal}}$ and $d_{\text{p}}$ to calculate $\tau_{\text{p}}$,
\begin{equation} \label{eq:tau_exp_scaling}
  \tau_{\text{p}} = \begin{cases}
  R_{\text{focal}} \sqrt{m_{\text{i}} / \left( T_{\text{e}} \bar{Z} \right)} & \text{for}\,\, R_{\text{focal}}<d_{\text{p}}\, , \\
  d_{\text{p}} \sqrt{m_{\text{i}} / \left( T_{\text{e}} \bar{Z} \right)} & \text{for}\,\, d_{\text{p}}\le R_{\text{focal}}\, .
  \end{cases}
\end{equation}
%

\section{Numerical results for spherical plasmas}
\label{sec:spherical}

\subsection{NEEC results}
\label{sec:spherical_neec}

The net NEEC rate $\lambda_{\text{neec}}$ is a function of the prevailing plasma conditions, e.g. electron temperature $T_{\text{e}}$ and density $n_{\text{e}}$. As presented in the previous Section, our model essentially consists of two separate parts which are combined to calculate the NEEC rate in the plasma: (i) the microscopic NEEC cross sections; (ii) the macroscopic plasma conditions like charge state distribution and electron flux.

The microscopic NEEC cross sections are calculated by employing bound atomic wave functions from a Multi-Configurational-Dirac-Fock method implemented in the GRASP92 package \cite{Grasp92} and  solutions of the Dirac equation with $Z_{\text{eff}}=q$ for the continuum. For both types of wavefunctions, we do not consider the effects of the plasma temperature and density, which are sufficiently small to be neglected in the final result of the nuclear excitation. The occurring nuclear matrix elements can be related to the reduced transition probability $B(E2)$ for which the calculated value of 3.5 W.u. (Weisskopf units) \cite{Hasegawa2011.PLB} was used.

\begin{figure}
    \includegraphics[width=\linewidth]{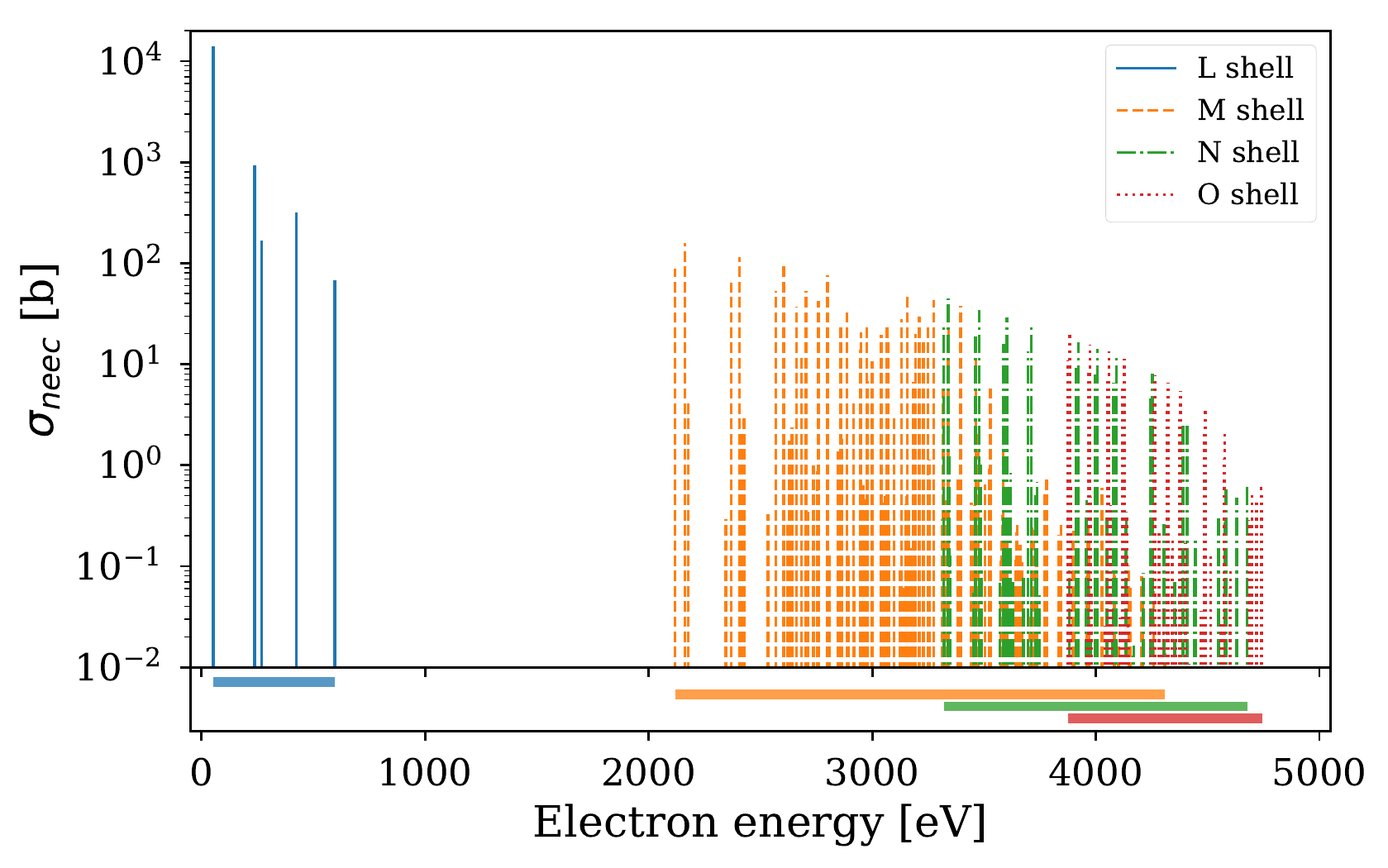}
    \caption{Microscopic NEEC cross sections as a function of the kinetic electron energy. The considered resonance channels for capture into the $L$-shell (blue, solid), $M$-shell (orange, dashed), $N$-shell (green, dash-dotted) and $O$-shell (red, dotted)  are shown together with the corresponding resonance energy windows (horizontal bars just above the x-axis).}
    \label{fig:sigma_neec}
\end{figure}

Numerical results for the individual NEEC cross sections of all considered capture channels are presented as a function of the kinetic electron energy in Fig.~\ref{fig:sigma_neec}.
We recall that the capture energy needs to coincide with the nuclear transition energy $E_{\text{n}}$, e.g. 4.85 keV in the case of ${}^{93\text{m}}$Mo triggering.
Therefore each peak represents a Lorentzian resonance located at $E_{\text{res}}=E_{\text{n}}-E_{\alpha_{\text{d}}}$ where $E_{\alpha_{\text{d}}}$ stands for the atomic binding energy of the considered capture channel $\alpha_{\text{d}}$. The width of the Lorentzian profile is given by the natural width of the nuclear triggering state $T$ which is composed of the radiative decay and internal conversion channels leading to approx. 130 neV.

Fig.~\ref{fig:sigma_neec} shows that NEEC prefers the capture into deeply bound states as  the cross section peak values increase for decreasing resonance energy of the free electrons. The resonance window for $L$-shell capture lies between 52 and 597 eV, for the $M$ shell between 2118 and 4308 eV, for the $N$ shell between 3320 and 4677 eV, and for the $O$ shell between 3874 and 4743 eV as illustrated by the horizontal lines in Fig.~\ref{fig:sigma_neec}.
Hence, there is a gap between approx. 600 and 2100 eV without any NEEC resonance channels.
Note that NEEC into the $K$ shell is energetically forbidden for the 4.85 keV transition in Mo.

The capture into the $2p_{3/2}$ orbital for ions with initial charge state $q=36$ and an initial electron configuration of $1s^2 2s^2 2p_{1/2}^2$ leads to the highest NEEC resonance strength (integrated cross section) of $2.78 \times 10^{-3}$ b eV. The corresponding resonance energy is at an electron energy of 52 eV. Interestingly, for higher charge states NEEC into the $L$ shell is energetically forbidden because the binding energies exceed the nuclear transition energy of 4.85 keV.

For the calculation of the net NEEC rate in the plasma, the microscopic NEEC cross sections have to be combined with the macroscopic plasma parameters according to Eqs.~\eqref{eq:neec-plasma.total} and \eqref{eq:neec-plasma.partial}.
We model the plasma conditions by a relativistic distribution for the free electrons and the converged charge state distribution computed with the radiative-collisional code FLYCHK \cite{FLYCHK2005} assuming the plasma to be in its non-local thermodynamical equilibrium (LTE) steady state.
The population kinetics model implemented in FLYCHK is based on rate equations including radiative and collisional processes, Auger decay and electron capture. These rate equations are solved for a finite set of atomic levels which consists of ground states, single excited states ($n\le$10), autoionizing doubly excited states and inner shell excited states for all possible ionic stages.
Employing a schematic atomic structure, the atomic energy levels are computed from ionization potentials where the effect of ionization potential depression occurring in plasmas is taken into account by the model of Stewart and Pyatt \cite{StewartAPJ1966}.

While the charge state distribution is described as a function of plasma parameters, $P_{q}(T_{\text{e}}, n_{\text{e}})$, a simplified model is used for the atomic orbital population.
For a specific charge state $q$, we assume (not necessarily to our advantage) that the charged ion is in its ground state initially (before NEEC) and capture of an additional electron occurs in a free orbital.

\begin{figure*}
    \includegraphics[width=\linewidth]{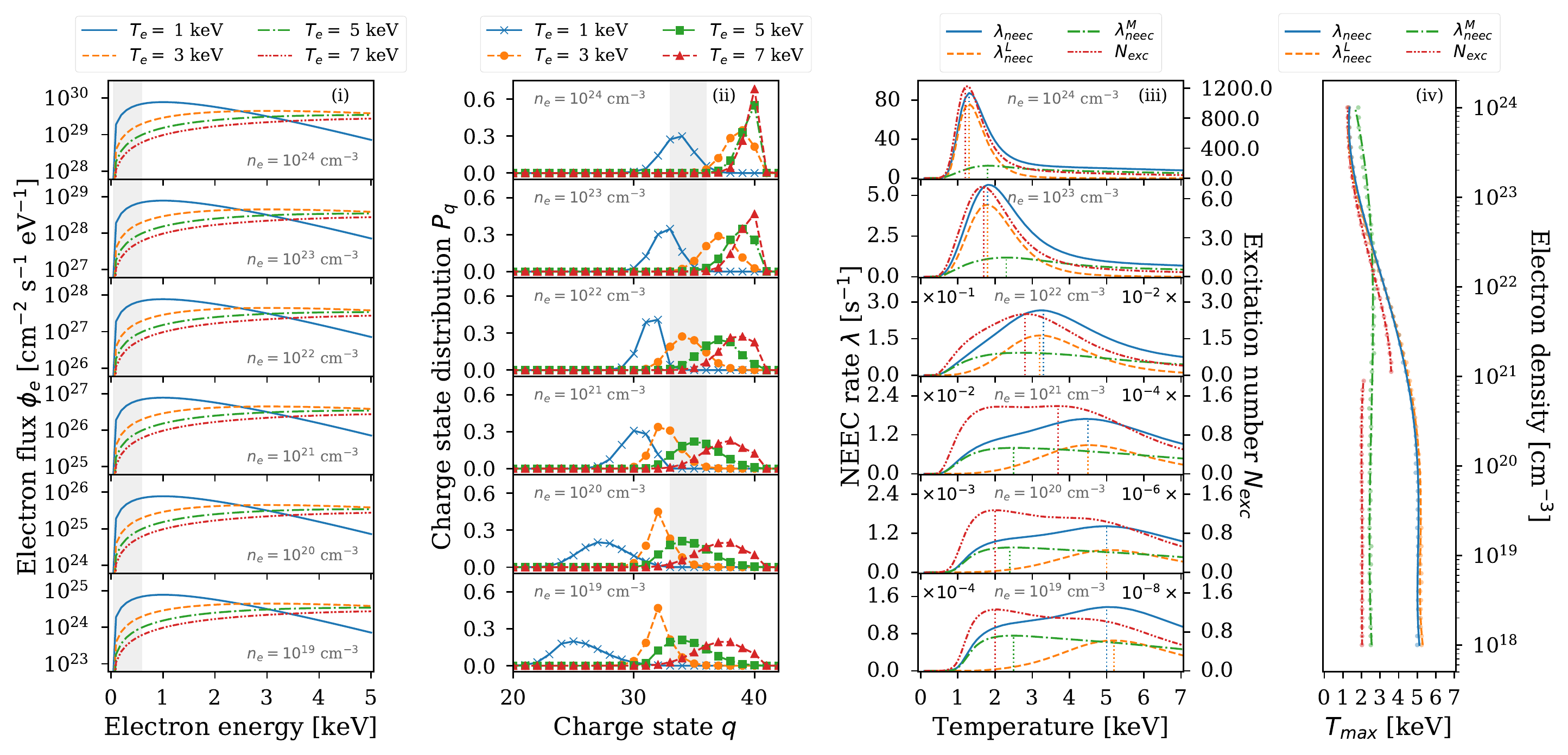}
    \caption{ From left to right:
    (i) Electron flux $\phi_{\text{e}}$ and (ii) charge state distributions (calculated with the help of FLYCHK) are shown for temperatures $T_{\text{e}} = 1$ keV (blue, solid curve for $\phi_{\text{e}}$, blue, solid curve with crosses for charge state distribution), 3 keV (orange, dashed curve for $\phi_{\text{e}}$, orange, dashed curve with filled circles for charge state distribution), 5 keV (green, dash-dotted curve for $\phi_{\text{e}}$, green, dash-dotted curve with filled squares for charge state distribution) and 7 keV (red, dash-dot-dotted curve for $\phi_{\text{e}}$, red, dashed curve with filled triangles for charge state distribution) at selected electron densities ranging from $10^{19}$ up to $10^{24}$ cm$^{-3}$. Electron energy and charge state ranges for $L$-shell capture are shaded in grey.
    (iii) NEEC rate $\lambda_{\text{neec}}$ (blue, solid curve) and the total number of excited isomers $N_{\text{exc}}$ (red, dash-dot-dotted curve), as well as the individual contributions $\lambda_{\text{neec}}^{\text{L}}$ (orange, dashed curve) and $\lambda_{\text{neec}}^{\text{M}}$ (green, dash-dotted curve) from the $L$ and $M$ shell, respectively, are shown as a function of the electron temperature $T_{\text{e}}$ for the selected electron densities $n_{\text{e}}$.
    A plasma radius of 40 $\mu$m has been assumed in the calculations of $N_{\text{exc}}$.
    (iv) The temperatures $T_{\text{max}}$ as function of density, for maximizing $N_{\text{exc}}$, $\lambda_{\text{neec}}$, $\lambda_{\text{neec}}^{\text{L}}$ and $\lambda_{\text{neec}}^{\text{M}}$, respectively, at each particular $n_{\text{e}}$.}
    \label{fig:general}
\end{figure*}

Numerical results for $\lambda_{\text{neec}}$ and the total number of excited isomers $N_{\text{exc}}$ are presented in Fig.~\ref{fig:general} in parallel with the corresponding electron fluxes and charge state distributions.
The calculation of the net NEEC rate involves charge states from $q=14$ up to the bare nucleus ($q=42$) with  333 NEEC capture states in total, composed of 5 $L$-shell, 168 $M$-shell, 70 $N$-shell and 90 $O$-shell orbitals.
The results for the dominant recombination channels  into the $L$ and $M$ atomic shells are presented individually in Fig.~\ref{fig:general}.
For the total NEEC rate $\lambda_{\text{neec}}$, further smaller contributions from the recombination into the $N$ and $O$ shells were also taken into account.
For the computation of $N_{\text{exc}}$ an arbitrary plasma radius of 40 $\mu$m has been assumed.

Both  $\lambda_{\text{neec}}$ and $N_{\text{exc}}$ increase with increasing electron density $n_{\text{e}}$.
In the range $n_{\text{e}} = 10^{19}$ cm$^{-3}$ to $10^{20}$ cm$^{-3}$, our calculations show that the charge state distribution $P_q$ is nearly unaffected for high temperatures $T_{\text{e}}$. In the same time, $\lambda_{\text{neec}}$ is enhanced by a factor of 10 maintaining almost the same functional dependence on electron temperature.
This  indicates that at low densities the boost in $\lambda_{\text{neec}}$ is (almost) a pure density effect coming from the increasing number of free electrons present in the plasma ($\phi_{\text{e}} \propto n_{\text{e}}$).
Increasing the electron density to even higher values, the behavior of $\lambda_{\text{neec}}$ and $N_{\text{exc}}$ becomes more involved as the charge distribution $P_q$ shows a complex dependence on the plasma conditions $n_{\text{e}}$ and $T_{\text{e}}$.
Between $n_{\text{e}} = 10^{21}$ cm$^{-3}$ and $10^{24}$ cm$^{-3}$ we see that  with increasing $n_{\text{e}}$ the atomic shell contributions change significantly and $\lambda_{\text{neec}}$ is substantially  enhanced.

Apart from the available charge states in the plasma, the match between the electron distribution [Fig.~\ref{fig:general}(i)] and the NEEC resonance conditions (Fig.~\ref{fig:sigma_neec}) plays an important role for this behavior.
The electron distributions reach their maxima at an energy $E \sim T_{\text{e}}$. For energies below this value (e.g. where the resonance energies for captures into the $L$ shell are located) more electrons are available for lower temperatures. In contrast, the high energy tail of the electron distribution drops exponentially with $\text{e}^{-E/ T_{\text{e}}}$, and is therefore faster decreasing the lower the temperature. In the case of Mo triggering and temperatures in the keV range, the energy region $E \gtrsim  T_{\text{e}}$, is in particular important for NEEC into the higher shells $M$, $N$ and $O$.

As seen from Fig.~\ref{fig:sigma_neec}, the best case for NEEC would be to have the maximum of the electron distribution ($E \sim T_{\text{e}}$) located at the resonance channels with the highest cross sections (e.g. capture into the $L$ shell).
However, assuming that the ions are always in their ground states initially, this condition cannot be exactly fulfilled, because lower temperatures also lead to lower charge states present in the plasma such that the $L$-shell capture will be closed.
The corresponding electron energy window and charge state range for the $L$-shell resonances are highlighted by the grey-shaded areas in Fig.~\ref{fig:general}. 

These contradicting requirements for efficient  NEEC suggest that there is a temperature $T_{\text{max}}$ at which the plasma-mediated NEEC triggering reaches a maximum for each density value $n_{\text{e}}$.
The temperatures $T_{\text{max}}$ for $N_{\text{exc}}$, the total and partial shell contributions $\lambda_{\text{neec}}$ are depicted as a function of the electron density in Fig.~\ref{fig:general}(iii) and Fig.~\ref{fig:general}(iv).
Naively, one would expect that $T_{\text{max}}$ is approximately the same for $N_{\text{exc}}$ and for $\lambda_{\text{neec}}$. This is however only true at high densities starting from $10^{21}$ cm$^{-3}$.
According to our approximation in Eq.~(\ref{eq:tau}), the chosen plasma lifetime is $T_{\text{e}}$-dependent. In particular at low electron densities $\tau_{\text{p}}$ acts as a weighting function proportional to $(T_{\text{e}})^{-1/2}$ shifting the maximum of $N_{\text{exc}}$ to lower temperatures. The optimal plasma conditions for the total excitation number can thus drastically differ from the optimal conditions for $\lambda_{\text{neec}}$ in this model. We  note  that the arbitrary choice of $R_{\text{p}}$ only influences the absolute scale of the NEEC excitation number, not the position of $T_{\text{max}}$.

\subsection{Ionization potential depression}
\label{sec:spherical_IPD}

While the effect of plasma-induced ionization potential depression (IPD) is taken into account for the charge state distribution (included in FLYCHK), it is neglected in the calculation of the microscopic NEEC cross sections and hence in the NEEC resonance energies so far.
In order to quantify the effect of the variation of atomic orbital energy on our final results, we adopted the model of Stewart and Pyatt  \cite{StewartAPJ1966} under the following assumptions:
(i) the bound electronic wave functions are unchanged;
(ii) the binding energies of atomic orbitals are lowered due the ionization potential depression, $\Delta V(q, T_{\text{e}}, n_{\text{e}})$;
(iii) the free-electron wavefunctions are computed with $Z_{\text{eff}}=q$;
(iv) the kinetic energy of the electrons required to match the NEEC resonance condition for a given orbital is modified according to the reduction of the corresponding binding energy.
Note that due to our approach for the potential lowering there might appear additional NEEC capture channels at low resonance energies (e.g. $L$ shell orbitals), while resonances disappear at resonance energies close to the nuclear transition energy.

The IPD given by the model of  Stewart and Pyatt \cite{StewartAPJ1966} is (using Gaussian units with $k_{\rm{B}}=1$)
\begin{equation} \label{eq:ipdwc}
  \Delta V = \frac{z e^2}{\lambda_{\rm{D}}}
\end{equation}
for the weak-coupling limit, i.e., $(a/\lambda_{\rm{D}})^3 \ll 1$ , and
\begin{equation} \label{eq:ipdsc}
  \Delta V = \frac{3}{2} \frac{z e^2}{a}
\end{equation}
for the strong-coupling limit, i.e., $(a/\lambda_{\rm{D}})^3 \gg 1$. Here, $\lambda_{\rm{D}}$ is the Debye length,
\begin{equation}
  \frac{1}{\lambda_{\rm{D}}^2} = \frac{4 \pi e^2}{T} (z^* + 1) n_{\rm{e}}\ ,
\end{equation}
and $a$ is the ion-sphere radius for an ion of net charge $z$ defined by
\begin{equation} \label{eq:ionsphere}
  4 \pi a^3 /3 = z/n_{\rm{e}}\ .
\end{equation}
The net charge $z$ is the charge state of the concerned ion (central ion) after the ionization, i.e., the above IPD expressions (\ref{eq:ipdwc})-(\ref{eq:ionsphere}) describe the process of removing an electron from an ion with charge $(z-1)^+$. Furthermore, $z^* = <\!\!z^2\!\!>\!\!/\!\!<\!\!z\!\!>$, $\!<\!\!z\!\!>$ is the averaged charge state of the plasma, and $<\!\!z^2\!\!>$ is the average of the charge state square of the plasma.

\begin{figure}
    \includegraphics[width=\linewidth]{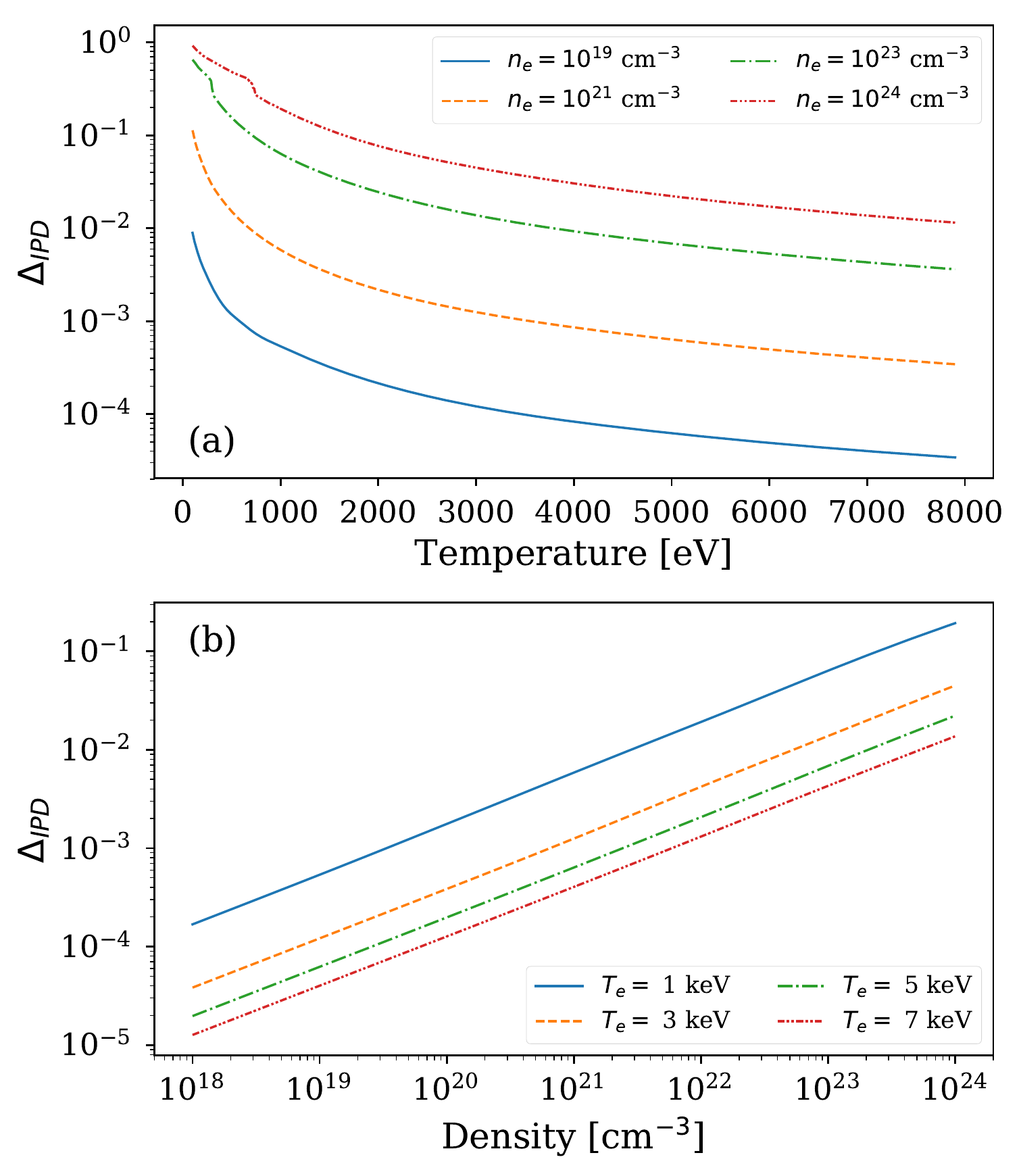}
    \caption{(a): the relative change in the NEEC rate due to IPD as a function of electron temperature for $n_{\text{e}}=10^{19}$ cm$^{-3}$ (blue, solid curve), $10^{21}$ cm$^{-3}$ (orange, dashed curve), $10^{23}$ cm$^{-3}$ (green, dash-dotted curve) and $10^{24}$ cm$^{-3}$ (red, dash-dot-dotted curve). (b): $\Delta_{\text{IPD}}$ as function of electron density for $T_{\text{e}}=1$ keV (blue, solid curve), 3 keV (orange, dashed curve), 5 keV (green, dash-dotted curve) and 7 keV (red, dash-dot-dotted curve). }
    \label{fig:IPD}
\end{figure}

The ionization potential lowering may change the binding energies from a few eV to hundreds of eV depending on the considered charge state $q$ and the prevailing plasma conditions in terms of $T_{\text{e}}$ and $n_{\text{e}}$.
In Fig.~\ref{fig:IPD} the relative change in the net NEEC rate due to IPD defined as
\begin{equation}
    \Delta_{\text{IPD}}(T_{\text{e}}, n_{\text{e}}) = \frac{\lambda_{\text{neec}}(T_{\text{e}}, n_{\text{e}}) - \lambda_{\text{neec}}^{\text{IPD}}(T_{\text{e}}, n_{\text{e}})}{ \lambda_{\text{neec}}(T_{\text{e}}, n_{\text{e}}) }
\end{equation}
is presented as a function of $T_{\text{e}}$ and $n_{\text{e}}$. The notation $\lambda_{\text{neec}}^{\text{IPD}}$ stands for the rate for which the plasma-induced potential lowering has been taken into account.

As seen from Fig.~\ref{fig:IPD}, the IPD leads to a decrease in the NEEC rate for all considered temperature and density values. The relative change becomes however only relevant for low temperatures and high densities. For instance, for $n_{\text{e}} \le 10^{19}$ cm$^{-3}$ or $T_{\text{e}} \ge 3$ keV, $\Delta_{\text{IPD}}$ stays below 1\% for the whole temperature or density range, respectively. Only for the extreme case of temperatures below 300 eV and densities higher than $10^{22}$ cm$^{-3}$, $\Delta_{\text{IPD}}$ reaches values of 10\% and above which is nevertheless still within the expected accuracy of our model. Deviations of these relative values  when employing other IPD models \cite{CiricostaPRL2012, HuPRL2017, LinPRE2017} than the one of Stewart and Pyatt also do not lead to a significant change of the NEEC results in the regime of interest.


\subsection{Resonant nuclear photoexcitation}
\label{sec:spherical_photo}

Nuclear excitation in the plasma may occur not only via NEEC but also via other mechanisms.
 In the considered temperature and density range the resonant nuclear photoexcitation is expected to be the main competing process to NEEC. For this reason we evaluated the photoexcitation rate in the plasma for two scenarios: (i) considering a black-body radiation spectrum; (ii) considering a photon distribution originating from bremsstrahlung. The theoretical expressions for the calculation have been given in Section \ref{sec:photoexc}.

\begin{figure}
    \includegraphics[width=1.0\linewidth]{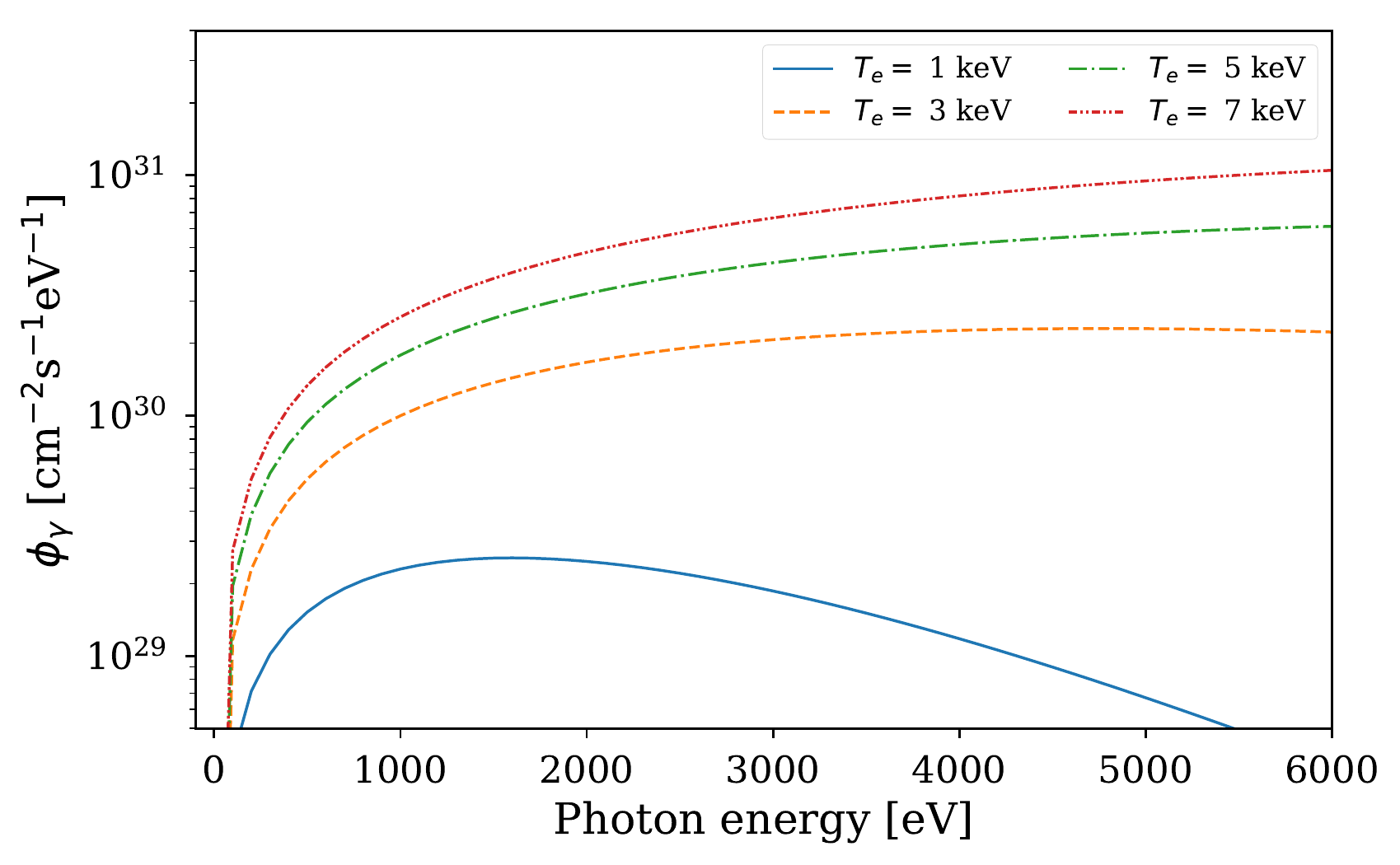}
    \caption{The blackbody photon flux $\phi_{\gamma}^{\text{TDE}}$ as a function of photon energy for temperatures $T_{\mathrm{e}}=1$ keV (blue, solid curve), 3 keV (orange, dashed curve), 5 keV (green, dash-dotted curve) and 7 keV (red, dash-dot-dotted curve).}
    \label{fig:photon_flux}
\end{figure}

The photon flux for TDE conditions (blackbody radiation) is presented in Fig.~\ref{fig:photon_flux}. In contrast to the electron flux, $\phi_{\gamma}^{\text{TDE}}$ is independent of the electron density and drastically rises with growing temperature $T_{\text{e}}$. For the ${}^{93\text{m}}$Mo isomer triggering especially the flux at  4.85 keV photon energy is interesting. The flux value increases from $7\times 10^{28}$ to $9\times 10^{30}$ cm$^{-2}$ s$^{-1}$ eV$^{-1}$
by going from $T_{\text{e}} = 1$ keV to $T_{\text{e}} = 7$ keV.

\begin{figure}
    \includegraphics[width=\linewidth]{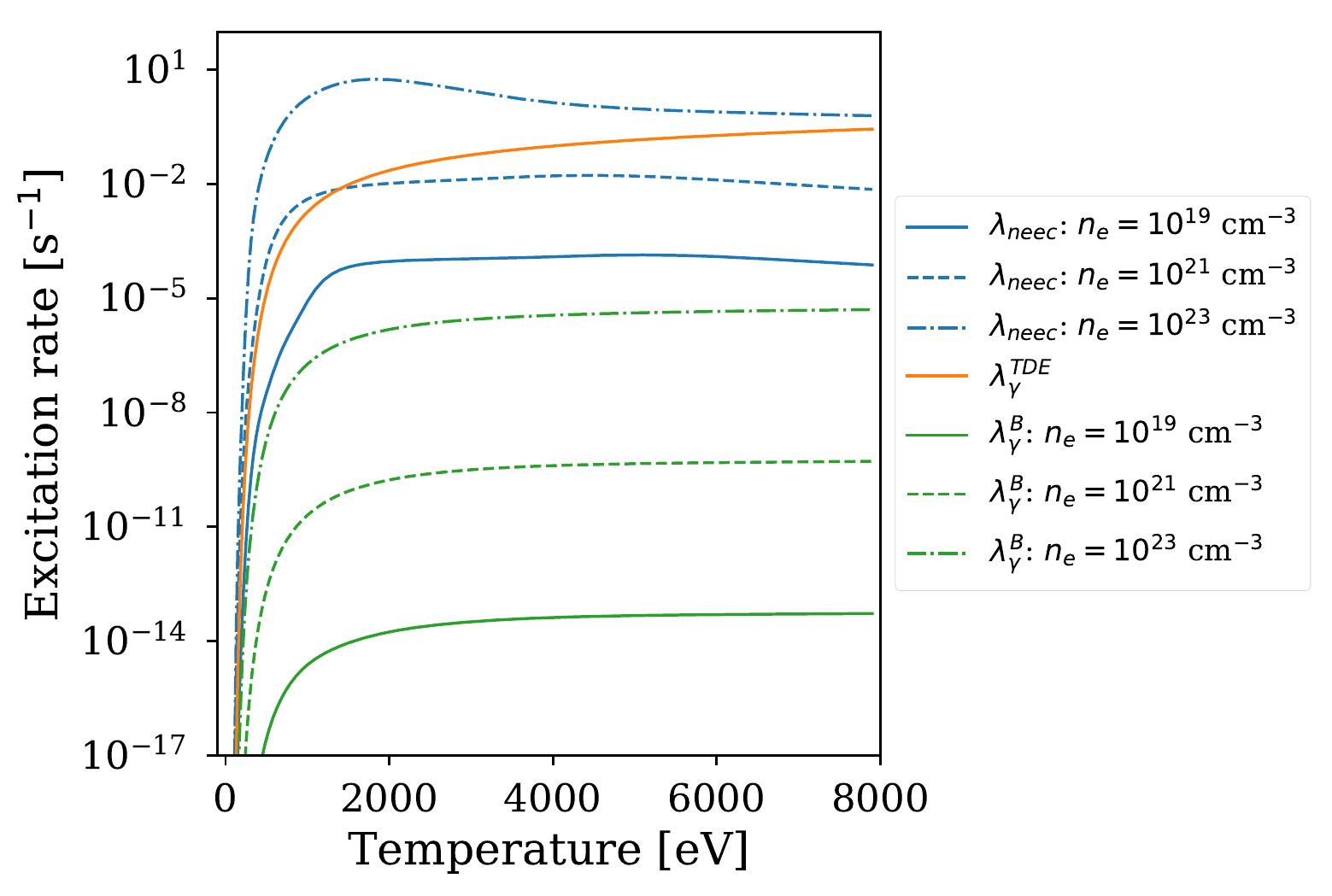}
    \caption{NEEC rate $\lambda_{\text{neec}}$ [blue (dark gray)] and photoexcitation rates $\lambda_{\gamma}^{\text{TDE}}$ [orange (light gray)] and $\lambda_{\gamma}^{\text{B}}$ [green (medium gray)] as functions of electron temperature. For NEEC and for photoexcitation via bremsstrahlung photons, electron densities of $10^{19}$ cm$^{-3}$ (solid lines), $10^{21}$ cm$^{-3}$ (dashed lines) and $10^{23}$ cm$^{-3}$ (dash-dotted lines) have been considered.}
    \label{fig:photoexc}
\end{figure}

In Fig.~\ref{fig:photoexc}, the NEEC rate $\lambda_{\text{neec}}$ is plotted together with the photoexcitation rates $\lambda_{\gamma}^{\text{TDE}}$ and $\lambda_{\gamma}^{\text{B}}$ for electron densities $10^{19}$, $10^{21}$ and $10^{23}$ cm$^{-3}$.
A comparison of the NEEC rate and the nuclear photoexcitation assuming a black-body radiation spectrum at the given plasma temperature $T_{\text{e}}$ shows that at  $n_{\text{e}} = 10^{21}$ cm$^{-3}$ NEEC dominates for $T_{\text{e}}\lesssim2$ keV and  for higher densities $n_{\text{e}} = 10^{22}$ cm$^{-3}$ up to a temperature of 6 keV. We note that while our NEEC values are to be considered as lower limit estimates, the actual photoexcitation in the plasma should be lower than the calculated values for a black-body spectrum in particular at low densities because photons may easier escape the finite plasma volume.
For the high density $n_{\text{e}} \ge 10^{23}$ cm$^{-3}$ parameter regime, NEEC is the dominant nuclear excitation mechanism.
The photoexcitation rate $\lambda_{\gamma}^{\text{B}}$ was calculated employing bremsstrahlung cross sections $d\sigma_{\text{B}}(E_{\text{e}}) / dE$ from Ref.~\cite{Pratt1977}. Our results show that the nuclear photoexcitation rate induced by bremsstrahlung photons is always several orders of magnitude lower and can be safely neglected.

\subsection{Plasma expansion: lifetime approach \& hydrodynamic model}
\begin{figure}
    \includegraphics[width=\linewidth]{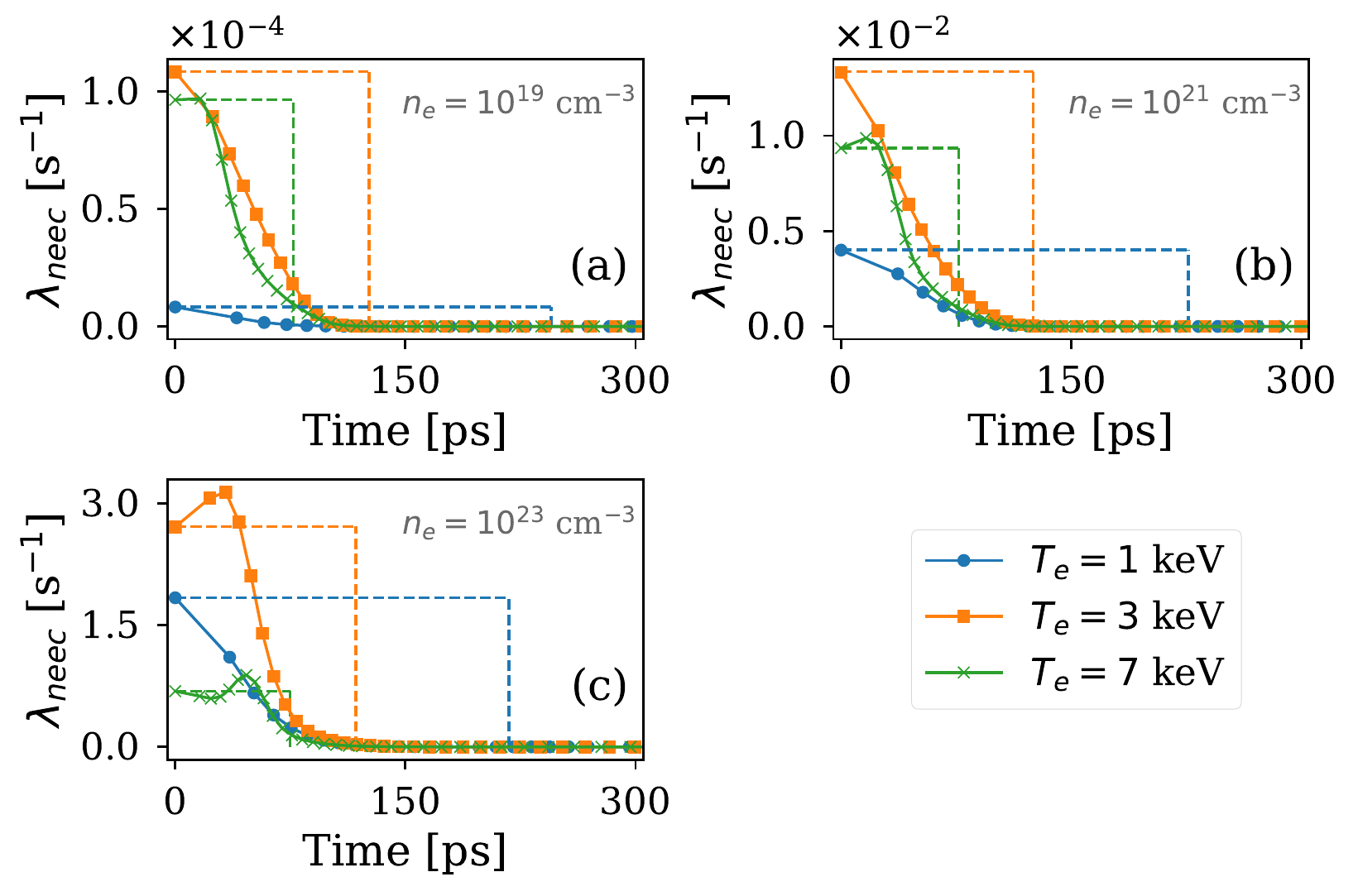}
    \caption{NEEC rate $\lambda_{\text{neec}}$ as a function of time during the hydrodynamic expansion. We considered initial electron densities $10^{19}$ cm$^{-3}$ (a), $10^{21}$ cm$^{-3}$ (b) and $10^{23}$ cm$^{-3}$ (c) and initial temperatures of 1 keV (blue curves with filled circles), 3 keV (orange curves with filled squares) and 7 keV (green curves with crosses). The lifetime estimate is illustrated by the dashed lines.}
    \label{fig:hydro_time}
\end{figure}

So far we considered the lifetime approach to estimate the total number of excited nuclei via Eq.~\eqref{eq:N_exc_simple}. For a more sophisticated ansatz, we apply the hydrodynamic model for the plasma expansion as introduced in Section \ref{sec:hydro} with initial conditions given by the present plasma conditions $n_{\text{i}}$, $T_{\text{e}}$ and $R_{\text{p}}$. In Fig.~\ref{fig:hydro_time}, the time evolution of $\lambda_{\text{neec}}$ during the expansion is presented for several initial plasma parameters.
As seen from the figure, the NEEC rate decreases over time since the plasma cools down and dilutes in terms of density while expanding. However, for cases where the initial temperature exceeds $T_{\text{max}}$ for the given initial density, the NEEC rate first increases, peaking out at optimal conditions and afterwards follows the typical decaying pattern, as clearly visible in Fig.~\ref{fig:hydro_time}.

A timescale comparison between the hydrodynamic model and the lifetime approach according to Eq.~\eqref{eq:tau} (illustrated by the dashed lines in Fig.~\ref{fig:hydro_time}) shows that the latter seems to overestimate the NEEC timescale especially for low temperatures. Also, the dependence of the NEEC timescale on $T_{\text{e}}$ is much weaker for the hydrodynamic expansion as determined by Eq.~\eqref{eq:tau}. For instance, for   $T_{\text{e}}=1$ keV and  $T_{\text{e}}=7$ keV at densities $n_{\text{e}} = 10^{21}$ cm$^{-3}$,
the lifetime $\tau_{\text{p}}$ is given by 226 and 77 ps, respectively. Considering the hydrodynamic expansion, the time integration over $\lambda_{\text{neec}}$ roughly converges after 110 and 130 ps, respectively.

\begin{figure}
    \includegraphics[width=0.8\linewidth]{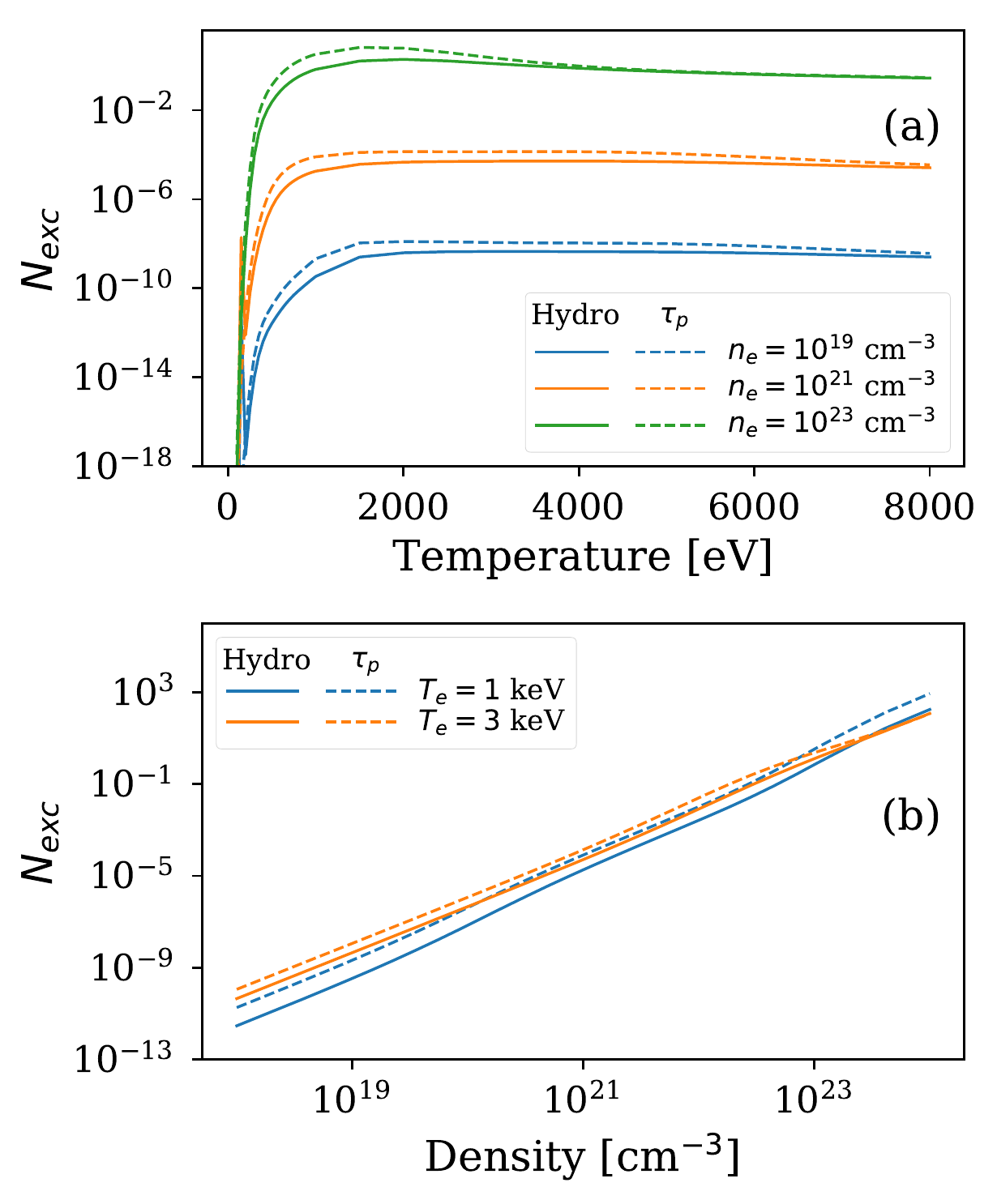}
    \caption{NEEC excitation number as a function of temperature (a) and electron density (b). Results from hydrodynamic expansion (solid curves) are shown in comparison to the lifetime estimate $\tau_{\text{p}}$ (dashed curves) for several initial plasma conditions. An initial plasma radius of 40 $\mu$m has been considered.}
    \label{fig:hydro_exc}
\end{figure}

Despite this discrepancy in the NEEC timescales between the lifetime and hydrodynamic models, the comparison of the total number of excited nuclei  as a function of the plasma conditions  in Fig.~\ref{fig:hydro_exc} shows a strickingly similar behaviour for the two models. For the calculations a plasma radius of 40 $\mu$m has been used. As a rule, the excitation numbers for the hydrodynamic expansion are slightly smaller than the corresponding ones from the lifetime approach. Furthermore, the highest deviation between the expansion models is at low temperatures and small densities as already expected from the results in Fig.~\ref{fig:hydro_time}. At a temperature of 1 keV and an electron density of $10^{19}$ cm$^{-3}$ the relative difference in the estimated excitation numbers $N_{\text{exc}}$ evaluates to 84\%, while it is on the order of 5\% at the high temperature ($\ge 6$ keV) and high density ($10^{23}$ cm$^{-3}$) tail.

Overall, the lifetime approach appears to provide reasonable estimates for $N_{\text{exc}}$ which deviate from the hydrodynamic model by 40-80\% for low temperatures and small densities. At high $T_{\text{e}}$ and $n_{\text{e}}$, the predicted excitation numbers almost coincide. As an advantage, the plasma lifetime approach is less computationally expensive and can be applied to a broader range of problems in comparison to the hydrodynamic expansion. For an order of magnitude estimate we therefore proceed to extract the optimal NEEC parameters for optical-laser generated plasmas with the help of the plasma lifetime  $\tau_{\text{p}}$ approach.

We note that, following the argument in Ref.~\cite{KrainovS2002}, we assume for Eq.~(\ref{eq:tionv}) that all ions have the same velocity $dR_{\rm{t}}/dt$ during the expansion. If one assumes the ion velocity at position $r$ to be $dr/dt$ and the velocity scales linearly with the position, the factor $3$ in Eq.~(\ref{eq:expaneq}) should be replaced by $5$  \cite{GuptaPRE2004, HilseLP2009}. However, the difference between these two factors should not affect our conclusion on the validity of the lifetime approach for the nuclear excitation calculation. Finally, the expansion with uniform density adopted above is a rather simplified model but it provides good estimates of the cluster expansion characteristics. For a more accurate model, however beyond the scope of the present work, we refer the interested reader to Ref.~\cite{GaoJAP2013}.

\section{Laser plasmas}
\label{sec:laser}

In the following we proceed to determine how the optimal NEEC parameter region in the temperature-density landscape may be accessed by a short laser pulse.  We discern in our treatment two cases, namely the low- and high-density plasmas, and refine accordingly our plasma model.

\subsection{Low density}
\label{sec:low}
First, we consider the case of a low-density (underdense) plasma, which can be generated via the interaction of a strong optical laser with a thin target. The plasma generation process typically evolves in two steps \cite{FuchsNP2006}: (i) a preplasma is formed by the prepulse of the laser; (ii) this preplasma is subsequently heated by the main laser pulse potentially up to keV electron energies.

We model the plasma following the approach presented in Section \ref{sec:plasma-scaling} with the help of  so-called scaling laws which provide a unique relation between laser parameters and plasma conditions. We employ two scaling law models, the sharp-edge scaling law (ponderomotive scaling law) in Eq.~(\ref{eq:SL1}) further denoted as SL1, and the short-scale length profile scaling law in Eq.~(\ref{eq:SL2}) referred to in the following as SL2.

\subsubsection{Results of scaling laws}

\begin{figure*}
    \includegraphics[width=\linewidth]{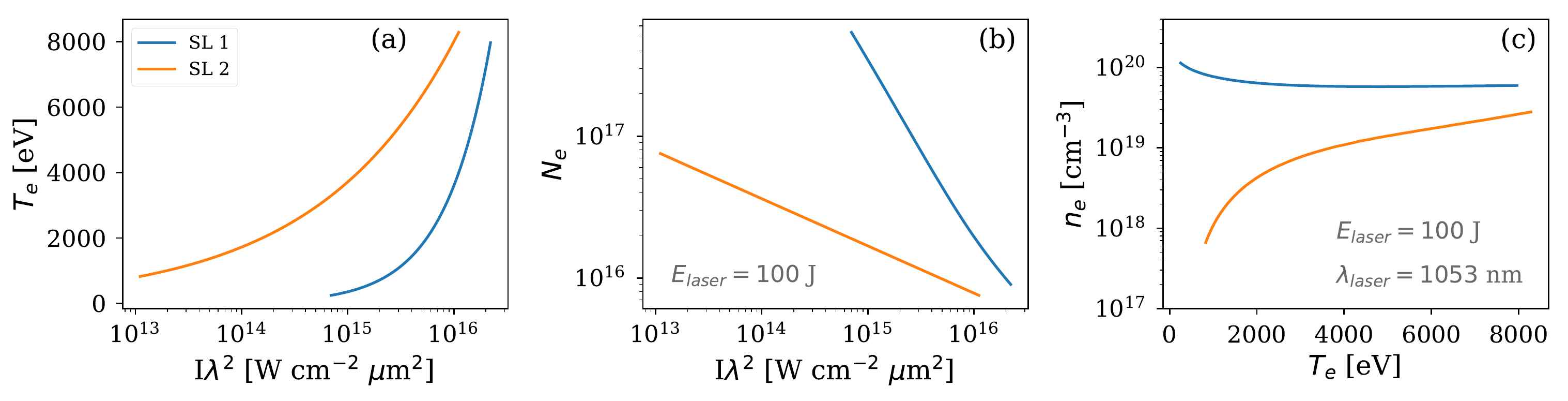}
    \caption{Accessible plasma parameters for ponderomotive scaling law [SL1, blue (dark gray) curves] and short-scale length profile scaling law [SL2, orange (light gray) curves]. The electron temperature (a) and the number of free electrons present in the plasma (b) are presented as functions of the laser irradiance $I \lambda^2$. In the calculations of $N_{\text{e}}$ a pulse energy of 100 J is assumed. The graph (c) shows the corresponding plasma profile $n_{\text{e}}=f(T_{\text{e}})$ for a laser wavelength of 1053 nm, a pulse energy of 100 J and intensity ranges $7\times10^{14} - 2\times10^{16}$ W/cm$^2$ (SL1) and  $10^{13} - 10^{16}$ W/cm$^2$ (SL2). The results here are independent of $\tau_{\text{laser}}$.}
    \label{fig:SL-profiles}
\end{figure*}

In Fig.~\ref{fig:SL-profiles}(a) and Fig.~\ref{fig:SL-profiles}(b) the electron temperature $T_{\text{e}}$ and the number of free electrons $N_{\text{e}}$ in the plasma are shown as functions of the laser irradiance $I \lambda^2$, respectively, for  both scaling laws SL1 and SL2. The considered range of irradiances has been chosen such that the expected electron temperatures span from approx. 300 eV to 8 keV.

SL1 and SL2 should be valid for collisionless heating in the intermediate irradiance regime $I \lambda^2 \sim 10^{16}$ W cm$^{-2}$ $\mu$m$^2$ \cite{GibbonPPCF1996}. In particular, SL1 covers the entire parameter regime presented in Fig.~\ref{fig:SL-profiles}, while SL2 exceeds its validity domain for the low-temperature case. For the sake of comparison, in the following we  compare SL1 and SL2 throughout the entire parameter regime of interest. However, we should keep in mind that at low temperatures (lower than $\sim 1$ keV) we enter an intensity regime lower than the lower limit for SL2 ($I \lambda^2 \sim 10^{14}$ W cm$^{-2}$ $\mu$m$^2$) \cite{GibbonPRL1992, GibbonPPCF1996}, in which the collisional heating may be the dominating heating mechanism.

In order to estimate the number of free electrons in the laser-heated plasma the laser absorption coefficient $f$ occuring in Eq.~\eqref{eq:absorption} needs to be fixed.
In Refs.~\cite{PingPRL2014, PricePRL1995} experimental data on the absorption of short laser pulses in the ultrarelativistic regime are presented. A Ti:sapphire laser with 150 fs pulses at 800 nm and an energy up to 20 J was used to heat Al foils (thickness $\sim 1.5 - 100\ \mu$m) and Si plates (thickness $\sim 400\ \mu$m). The measured laser absorption shows no significant dependence on the target thickness and the material. Moreover, in consistency with previous experiments at lower intensities \cite{PricePRL1995}, it could be shown that the absorption mechanisms change from collision dominated to collisionless by exceeding an intensity of around $10^{17}$ W/cm$^2$. The experimental results show a good agreement with a theoretical calculation based on a Vlasov-Fokker-Planck code.
We therefore adopt a universal absorption coefficient $f=f(I \lambda^2)$ as a function of laser irradiance to estimate the absorbed laser energy by peforming a cubic interpolation to theoretical results based on a Vlasov-Fokker-Planck code presented in Ref.~\cite{PingPRL2014}. A more detailed discussion on the laser absorption coefficient and its impact on the NEEC rates is given in Section \ref{fabs}.

As can be seen from Fig.~\ref{fig:SL-profiles}, SL2 predicts higher temperatures for a given laser intensity $I_{\text{laser}}$. However, since the number of free electrons is inversely proportional to the electron temperature [see Eq.~\eqref{eq:absorption}], the resulting density for SL2 is expected to be smaller in comparison to SL1.
For a wavelength of $\lambda_{\text{laser}} = 1053$ nm typical for Nd:glass lasers and a pulse energy of 100 J, the corresponding temperature-density profiles for SL1 and SL2 are presented in Fig.~\ref{fig:SL-profiles}(c).
With the considered intensity range between $7\times10^{14}$ and $2\times10^{16}$ W/cm$^2$ for SL1, the absorption fraction $f$ lies between 0.1 and 0.2 leading to electron densities in the order of $10^{20}$ cm$^{-3}$. The extension of the low intensity tail for SL2 ($10^{13} - 10^{16}$ W/cm$^2$) results in slightly higher absorption coefficients up to 30\%. However, the electron densities are smaller in this case since the average electron temperature is higher for a fixed plasma volume $V_{\text{p}}$.

\begin{figure}
    \includegraphics[width=\linewidth]{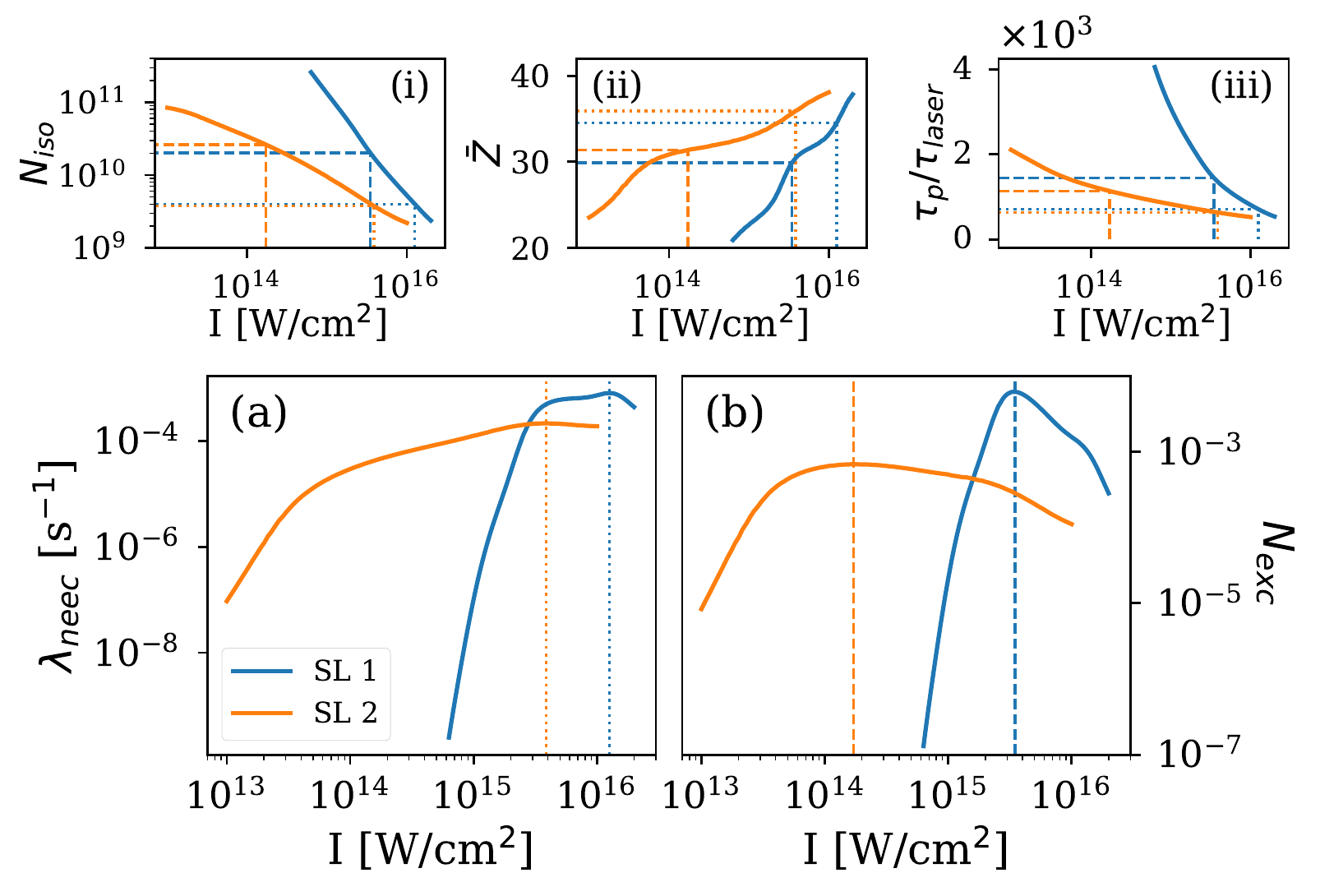}
    \caption{The NEEC rate $\lambda_{\text{neec}}$ (a) and the total excitation number $N_{\text{exc}}$ per laser pulse (b) as functions of laser intensity for ponderomotive scaling law [SL1, blue (dark gray) curves] and short-scale length profile scaling law [SL2, orange (light gray) curves]. Upper graphs, from left to right: (i) the number of isomers present in the plasma, (ii) the average charge state $\bar{Z}$ and (iii) the plasma lifetime are presented. See text for further explanations.}
    \label{fig:low}
\end{figure}

Numerical  results for  $\lambda_{\text{neec}}$ and for the total excitation number $N_{\text{exc}}$ per laser pulse  are presented in
Fig.~\ref{fig:low} as functions of the laser intensity.
The plasma expansion time is estimated by using the lifetime approach with the smallest length scale out of $R_{\text{focal}}$ and $d_{\text{p}}$ [see Eq.~\eqref{eq:tau_exp_scaling}] for a lower-limit estimate of the NEEC excitation.
We consider a pulse energy of 100 J, wavelength of 1053 nm, and laser pulse duration values of 500 fs.
Apart from $\lambda_{\text{neec}}$ and $N_{\text{exc}}$, the number of isomers $N_{\text{iso}}$ present in the plasma, the average charge state $\bar{Z}$ and the plasma lifetime $\tau_{\text{p}}$ in units of $\tau_{\text{laser}}$ are shown as functions of the laser intensity.

Analog to the discrepancy between $T_{\text{max}}$ for NEEC rate and total excitation number, also here the optimal laser intensities $I_{\text{opt}}$ at which $\lambda_{\text{neec}}$ and respectively $N_{\text{exc}}$ are maximal do not coincide.
For the assumed laser parameters, $\lambda_{\text{neec}}$ is maximized by
$I_{\text{opt}} = 1.26\times 10^{16}$ W/cm$^2$ at a temperature of 5 keV and a density of $5.8\times 10^{19}$ cm$^{-3}$ in the case of SL1.
In contrast, the optimal intensity for $N_{\text{exc}}$ per laser pulse is  $3.5\times 10^{15}$ W/cm$^2$. The electron temperature and density achieved at this intensity are 1.4 keV and $7.0\times 10^{19}$ cm$^{-3}$, respectively, leading to a charge state distribution with $\bar{Z} \sim 30$ where capture channels into the $M$ shell still dominate the $L$-shell contribution.

The optimal values for SL2 lie at smaller intensities, $3.9\times 10^{15}$ W/cm$^2$ and $1.72\times 10^{14}$ W/cm$^2$ for $\lambda_{\text{neec}}$ and $N_{\text{exc}}$, respectively, where plasma conditions $T_{\text{e}} = 6$ keV, $n_{\text{e}} = 3.8\times10^{19}$ cm$^{-3}$ and
$T_{\text{e}} = 2.1$ keV, $n_{\text{e}} = 4.7\times10^{18}$ cm$^{-3}$ are prevailing.

In general, Fig.~\ref{fig:low} shows that the total number of excited isomers is maximal at plasma conditions with lower average charge state $\bar{Z}$ but longer plasma lifetime and larger plasma volume in comparison to the optimal conditions for the NEEC rate. The effect of the larger plasma volume can be nicely seen for the number of isomers present in the plasma [Fig.~\ref{fig:low}(i)] which is given by approximatively $2\times10^{10}$ isomers at $I_{\text{opt}}$ for $N_{\text{exc}}$. The proportionality of $N_{\text{iso}}$ with respect to the laser irradiance $I \lambda^2$ is given by the following relation,
\begin{equation}
    N_{\text{iso}} \propto \bar{Z} N_{\text{e}} \propto
    \begin{cases}
        \bar{Z} f(I \lambda^2) / (I \lambda^2) & \text{for SL1} \\
        \bar{Z} f(I \lambda^2) / (I \lambda^2)^{1/3} & \text{for SL2}
    \end{cases}\ ,
\end{equation}
where $\bar{Z}$ is itself a function of laser intensity and wavelength for fixed pulse energy $E_{\text{pulse}}$.

\subsubsection{Laser absorption \label{fabs}}
\begin{figure}
    \includegraphics[width=0.8\linewidth]{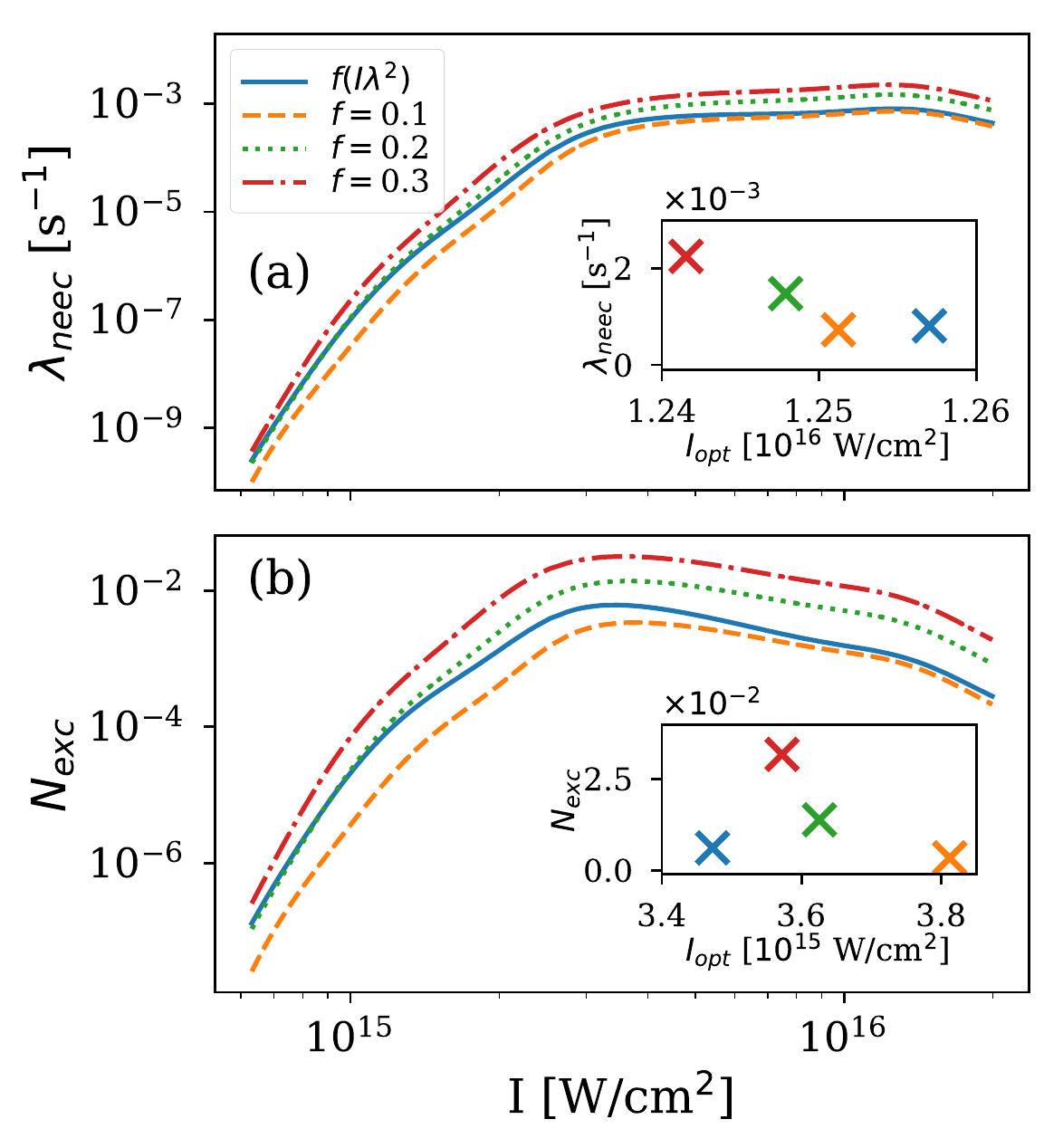}
    \caption{Impact of the laser absorption $f$ on NEEC results considering SL1. $\lambda_{\text{neec}}$ (a) and $N_{\text{exc}}$ (b) are presented as functions of $I_{\text{laser}}$ for $f=f(I\lambda^2)$ (blue, solid line), $f=0.1$ (orange, dashed line), $f=0.2$ (green, dotted line) and $f=0.3$ (red, dash-dotted line). The inset presents the optimal laser intensities for NEEC for the different $f$ values. See text for further explanations.}
    \label{fig:f_SL1}
\end{figure}

\begin{figure}
    \includegraphics[width=0.8\linewidth]{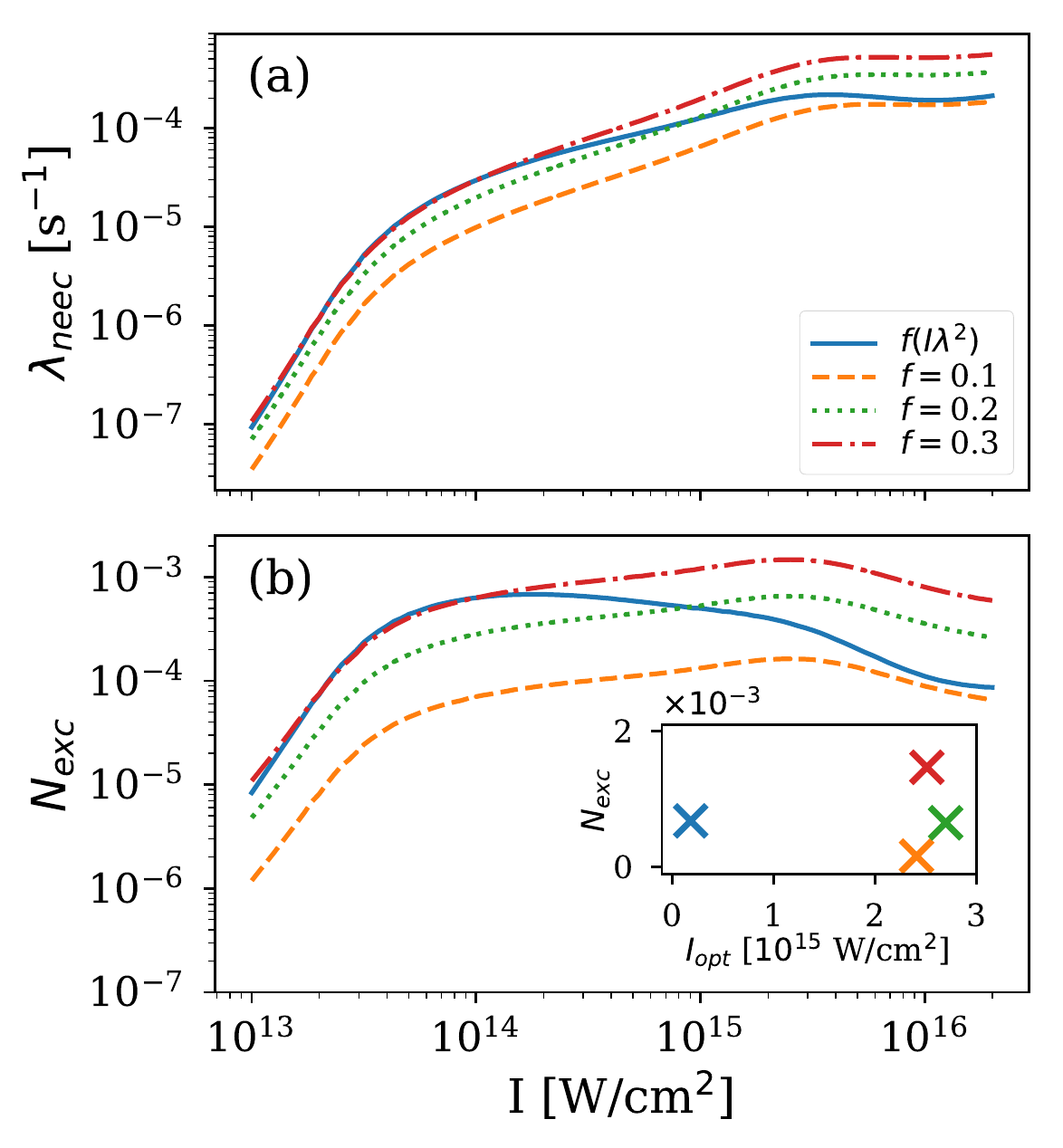}
    \caption{Impact of the laser absorption $f$ on NEEC results considering SL2. Analog notation as in Fig.~\ref{fig:f_SL1}.}
    \label{fig:f_SL2}
\end{figure}

Since the laser absorption fraction $f$ can be treated as free parameter in the scaling laws SL1 and SL2, we study the effect of different $f$ values for the  NEEC rate in the plasma and the corresponding nuclear excitation. We have performed calculations with constant absorption fractions 10\%, 20\% and 30\% considering a pulse energy of 100 J, a laser wavelength of 1053 nm and a pulse duration of 500 fs. The results together with a comparison with the laser absorption model $f(I \lambda^2)$ are shown in Fig.~\ref{fig:f_SL1} for SL1 and Fig.~\ref{fig:f_SL2} for SL2.


As can be seen from these two figures, the NEEC rate as well as the total excitation number increase with increasing absorption coefficient $f$ since higher densities are reached. Moreover, the optimal laser intensities  depend on the laser absorption as illustrated in the corresponding insets of the graphs. While the change of $I_{\text{opt}}$ due to $f$ is smaller than 10\% and hence negligible in terms of the expected accuracy of our model for SL1, the intensity and wavelength-dependent absorption coefficient $f$ has a much stronger effect on the predictions of SL2 in comparison to a constant absorption fraction.
With constant $f$ the optimal intensity for $N_{\text{exc}}$ is at approx. $2.5\times10^{15}$ W/cm$^2$ in comparison to $1.7\times10^{14}$ W/cm$^2$ with $f(I \lambda^2)$.
Note that  $I_{\text{opt}}$ for $\lambda_{\text{neec}}$ is not shown in Fig.~\ref{fig:f_SL2} since the rate keeps increasing for higher intensities way out of the validity range for the applied scaling law model SL2.

\subsubsection{Dependence on laser parameters}

\begin{figure}%
    \centering
    \includegraphics[width=1.0\linewidth]{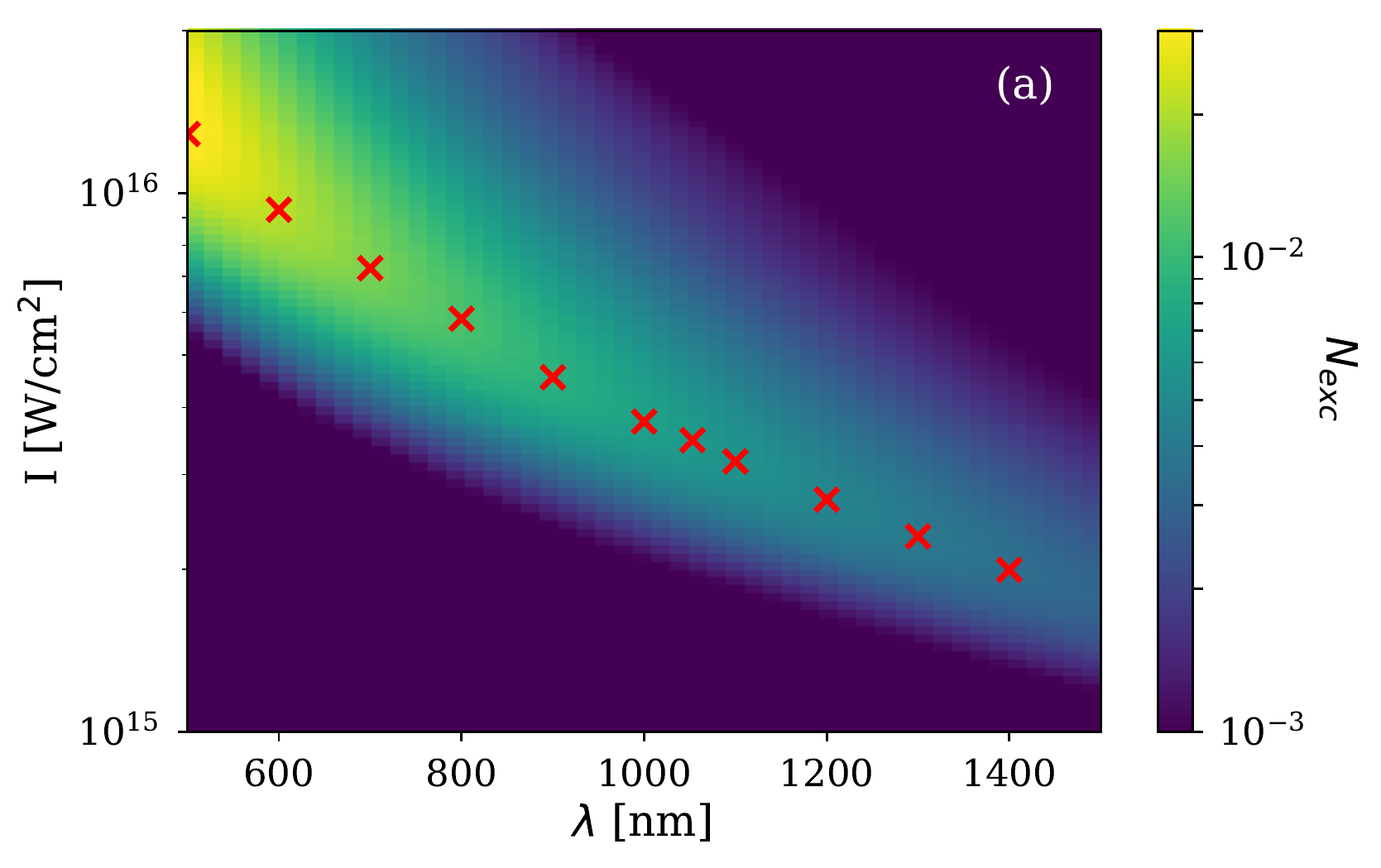}\\%
    \includegraphics[width=1.0\linewidth]{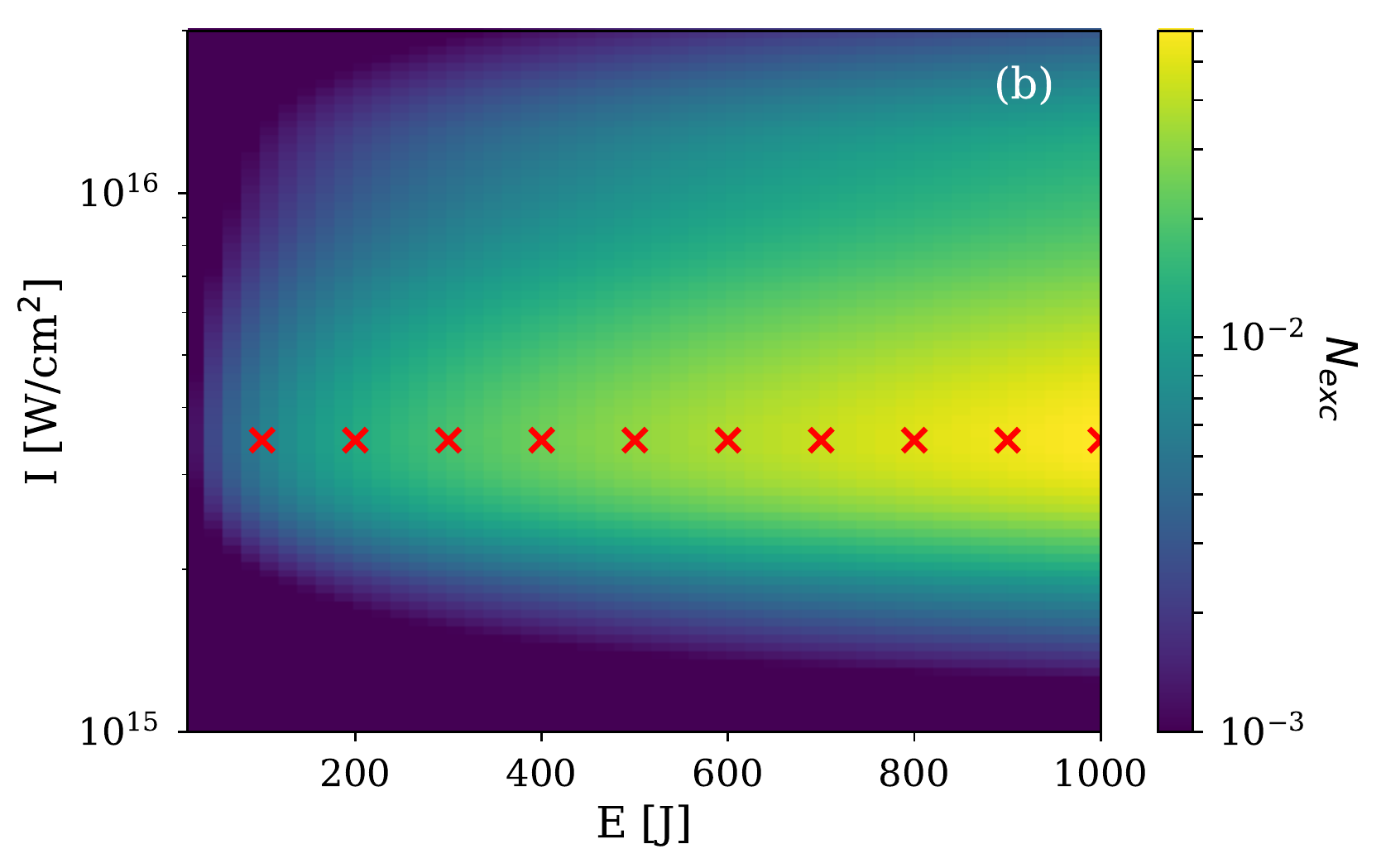}%
    \caption{Total number of excited isomers as a function of laser parameters. $N_{\text{exc}}(I,\lambda)$ for fixed $E_{\text{pulse}}=100$ J (a) and $N_{\text{exc}}(I,E)$ for fixed $\lambda_{\text{laser}}=1053$ nm (b) are presented. The pulse duration is assumed to be $\tau_{\text{laser}}=500$ fs.}
    \label{fig:SL_grid}
\end{figure}

In this Section, we analyze the functional behavior of the NEEC excitation on the laser parameters $I_{\text{laser}}$, $\lambda_{\text{laser}}$, $\tau_{\text{laser}}$ and $E_{\text{pulse}}$. For this analysis we restrict ourselves to the steep density gradients scenario which is best described by SL1 using the universal absorption coefficient $f(I \lambda^2)$. In Fig.~\ref{fig:SL_grid}(a) we present the total number of excited isomers $N_{\text{exc}}$ as a function of laser intensity and wavelength for fixed laser pulse duration  500 fs and pulse energy  100 J.
 It can be seen that the highest excitation numbers can be found for small wavelengths at the corresponding optimal intensity $I_{\text{opt}}$ illustrated by the red crosses for given $\lambda_{\text{laser}}$.
The optimal intensity values are increasing with decreasing laser wavelength. We recall that smaller wavelengths lead to smaller electron densities which require a higher temperature to maximize the NEEC excitation (cf. Fig.~\ref{fig:general}).

However, typically the wavelength is a parameter determined by the fundamental laser design (i.e. 1053 nm in the case of Nd:glass lasers, or 800 nm for Ti:sapphire lasers) and can only be changed by considering higher harmonics. Therefore it is worth to further investigate the behavior of $N_{\text{exc}}$ in terms of variable laser intensity and pulse energy for a given wavelength of, for instance 1053 nm, and pulse duration of 500 fs. Numerical results are shown in Fig.~\ref{fig:SL_grid}(b). As can be expected already from the direct relation between number of free electrons in the plasma and $E_{\text{pulse}}$ [compare Eq.~\eqref{eq:absorption}], the NEEC excitation is higher for higher pulse energies.
The optimal laser intensity for given $E_{\text{pulse}}$ [represented by the red crosses in Fig.~\ref{fig:SL_grid}(b)] is constant over the considered energy range.

\begin{figure}
    \includegraphics[width=\linewidth]{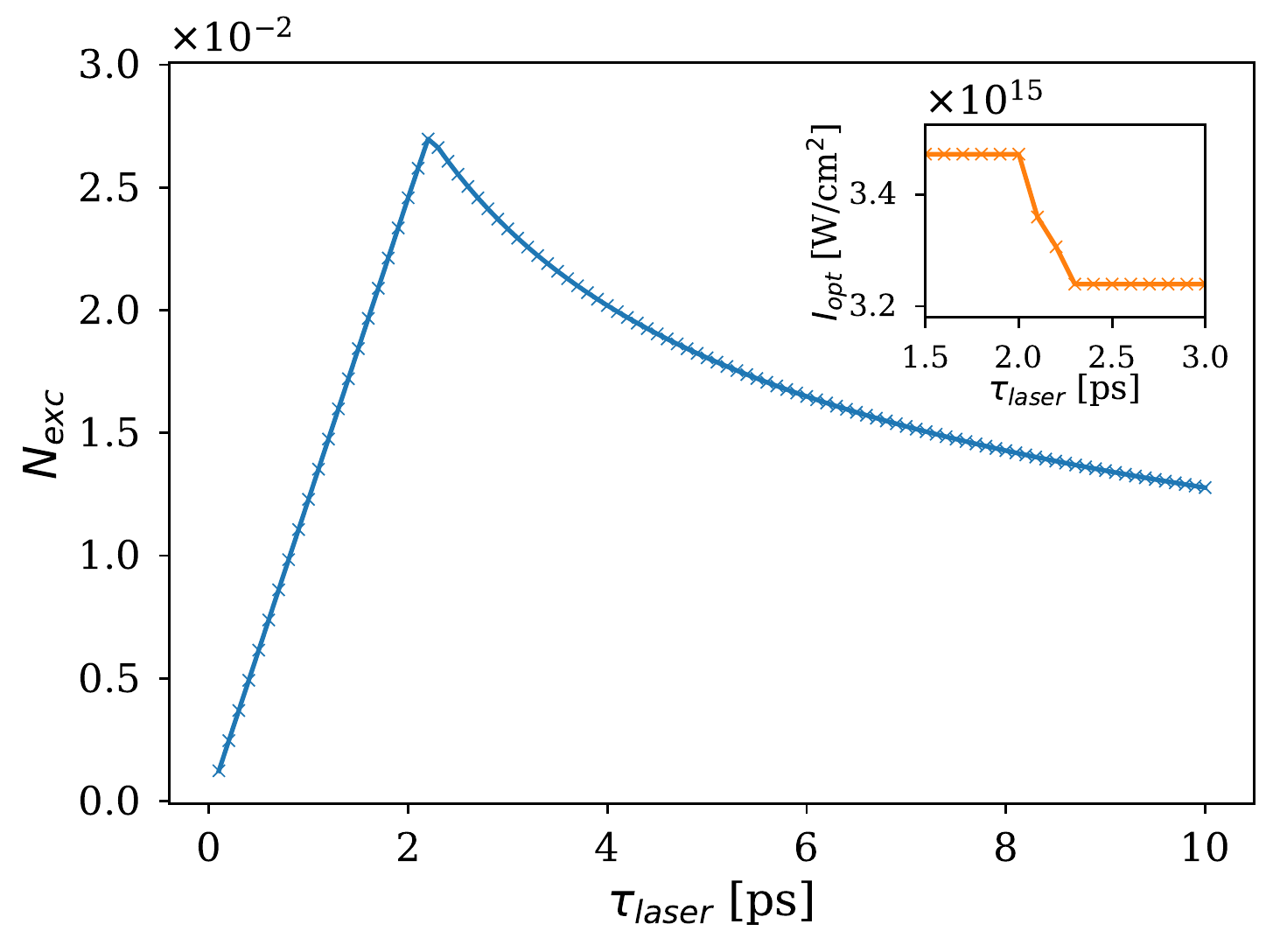}
    \caption{Total excitation number $N_{\text{exc}}$ at optimal intensity as a function of $\tau_{\text{laser}}$. The inset shows $I_{\text{opt}}(\tau_{\text{laser}})$ close to the region where $N_{\text{exc}}$ is maximized. We considered a pulse energy of 100 J and laser wavelength of 1053 nm.}
    \label{fig:SL_tau}
\end{figure}

For $d_{\text{p}}<R_{\text{focal}}$ (the case for the parameters of Figs.~\ref{fig:low} - \ref{fig:SL_grid}) the plasma lifetime is determined by $d_{\text{p}}$ and in turn by $\tau_{\text{pulse}}$.
In order to evaluate the influence of $\tau_{\text{pulse}}$ on the NEEC excitation, Fig.~\ref{fig:SL_tau} shows $N_{\text{exc}}$ at the optimal intensity $I_{\text{opt}}$ as a function of the pulse duration. For the calculations we considered again a fixed pulse energy of 100 J and a wavelength of 1053 nm.
As seen from the figure, the NEEC excitation becomes stronger with increasing laser pulse duration $\tau_{\text{pulse}}$ reaching its maximum at 2.2 ps, the value where $d_{\text{p}}=R_{\text{focal}}$. According to our model, this condition is satisfied for
\begin{equation}
    \tau_{\text{laser}} = \left( \frac{E_{\text{laser}}}{c^2 \pi I_{\text{laser}}} \right)^{1/3}\ .
\end{equation}
For even longer pulse durations we need to use  $R_{\text{focal}}$ in our model to determine the plasma lifetime [see Eq.~\eqref{eq:tau_exp_scaling}].
We then notice a decrease of $N_{\text{exc}}$ as for a given laser pulse energy, longer $\tau_{\text{pulse}}$ values require smaller focal radii to obtain the same intensity and in turn shorter plasma lifetime.
The optimal intensity shifts slightly to smaller values by going from the parameter region where $d_{\text{p}}<R_{\text{focal}}$ to parameters with $d_{\text{p}}>R_{\text{focal}}$ (see inset of Fig.~\ref{fig:SL_tau}).

We note that for short-pulse lasers, the plasma lifetime is constrained by the plasma thickness which depends on the pulse duration and optimally should have a size similar to the other two dimensions encompassed by the focal spot. We have not considered here long nanosecond laser pulses which lead to a complex plasma evolution which is difficult to model.
While short laser pulses limit the plasma lifetime, long, ns laser pulses in turn come with a small focal spot to obtain the necessary laser intensities. Effectively, the nuclear excitation should be similar in magnitude at a given pulse energy for both short-pulse and ns-pulse lasers.

\subsubsection{High-power laser facilities}

In this Section, we evaluate the optimal laser intensity $I_{\text{opt}}$ and  the expected maximal NEEC excitation $N_{\text{exc}}$ for realistic parameters of high-power optical lasers which are available currently or under construction. Results for the ELI-beamlines L4 \cite{ELI-BL-web}, ELI-NP \cite{NegoitaRRP2016, ELI-NP-web}, PETAL \cite{CasnerHEDP2015}, LULI \cite{LULI-web}, VULCAN \cite{Vulcan-web} and PHELIX \cite{PHELIX-web} lasers are presented in Table \ref{table:PWlasers}.

For all considered cases, the excitation $N_{\rm{exc}}$ per laser pulse is orders of magnitude larger than the one in the XFEL-generated cold plasma \cite{GunstPRL2014, GunstPOP2015}. We note that in the analysis in Refs. \cite{GunstPRL2014, GunstPOP2015}, a $B(E2)$ value of $1$ W.u. and the isomer fraction of $\sim 1.8 \times 10^{-7}$ were used. For the comparison presented here, we have recalculated the XFEL excitation number using $B(E2) = 3.5$ W.u. and the isomer fraction ($\sim 10^{-5}$)  considered for the present work yielding the result $N_{\rm{exc}} \sim 10^{-6}$ for a $T_{\rm{e}} = 350$ eV plasma.
Moreover, the largest value of $2.2$ excitations per pulse should be reached with the PETAL laser which provides both high laser power and longer pulse duration $\tau_{\text{pulse}}$. Table  \ref{table:PWlasers} and Figs.~\ref{fig:SL_grid} and \ref{fig:SL_tau} show that a balance between the laser power and laser pulse duration is beneficial for the excitation number.

Note that the values presented here slightly differ from the values provided in Ref.~\cite{Wu2018}, since we took into account an additional data point for the fitting of the universal absorption coefficient $f$ to extend the model to smaller irradiances $I\lambda^2$.

\renewcommand{\arraystretch}{1.25}
\begin{table*}
  \centering
  \footnotesize
  \begin{tabular}{lcccccccc}
  \hline\hline
  & & & & \tabularnewline[-0.4cm]
  & ELI-beamlines & ELI-NP & PETAL & LULI & VULCAN & PHELIX \tabularnewline
  & & & & \tabularnewline[-0.4cm] \hline
  & & & & \tabularnewline[-0.4cm]
  $E_{\text{pulse}}$ [J] &1500 & 250 & 3500 & 100 & 500 & 200 \tabularnewline
  $\tau_{\text{pulse}}$ [fs] &150 & 25 & 5000 & 1000 & 500 & 500 \tabularnewline
  $\lambda$ [nm] & 1053 & 800 & 1053 & 1053 & 1053 & 1053 \tabularnewline
  & & & & \tabularnewline[-0.4cm] \hline
  & & & & \tabularnewline[-0.4cm]
  $N_{\text{exc}}$ & $2.8\times10^{-2}$ & $1.4\times10^{-3}$ & $2.2$ & $1.2\times10^{-2}$ & $3.1\times10^{-2}$ & $1.2\times10^{-2}$ \tabularnewline
  & & & &  \tabularnewline[-0.4cm] \hline\hline
  \end{tabular}
  \caption{Laser parameters and maximal $N_{\text{exc}}$ achieved at the optimal laser intensity $I_{\text{opt}} = 3.4\times 10^{15}$ W/cm$^2$ for  ELI-beamlines L4 \cite{ELI-BL-web}, PETAL \cite{CasnerHEDP2015}, LULI \cite{LULI-web}, VULCAN \cite{Vulcan-web} and PHELIX \cite{PHELIX-web} and $I_{\text{opt}} = 5.7\times 10^{15}$ W/cm$^2$ for ELI-NP \cite{NegoitaRRP2016, ELI-NP-web} lasers. }
  \label{table:PWlasers}
\end{table*}
\renewcommand{\arraystretch}{1}


\subsection{High density}
\label{sec:high}

We now turn to the case of high electron densities, which promises the strongest nuclear excitation  according to Fig.~\ref{fig:general}. Experiments and simulations have shown that it is possible to isochorically heat targets at solid-state density to temperatures of a few hundred eV or even a few keV \cite{SaemannPRL1999, AudebertPRL2002, SentokuPOP2007}. Since in this regime the heating of the target is mainly conducted by secondary particles, i.e. hot electrons generated in the laser-target interaction, a  more sophisticated model is necessary  compared to the low-density case.

\subsubsection{PIC simulation}
The solid-state isomer target is practically a Niobium foil with a $10^{-5}$ fraction  of embedded ${}^{93\text{m}}$Mo isomers.
We have performed a one-dimensional (1D)  particle-in-cell (PIC) simulation of a Nb solid target with 1 $\mu$m thickness and Nb density of $n_{\text{nb}} = 5.5 \times 10^{22}$ cm$^{-3}$ interacting with a high-power laser using the EPOCH code \cite{ArberPPCF2015}. The isomer fraction  is small enough to be neglected here in the determination of the plasma conditions.
The laser has a Gaussian profile in time with peak intensity $I = 10^{18}$ W/cm$^2$,  laser duration  $\tau_{\text{pulse}} = 500$ fs, and laser wavelength $\lambda = 800$ nm, respectively. At the boundary of the simulation box where the laser is introduced, the laser reaches the peak intensity at time $t = 500$ fs.
A linear preplasma with the thickness of $0.5$ $\mu$m is considered in front of the solid target.  The simulation box is $4$ $\mu$m in length, and the solid target is placed at the center of the simulation box. Ionization is not included explicitely in the simulation;  as a representative  order for the electron density, we  fix the charge state to 10.

\begin{figure*}
    \includegraphics[width=0.83\linewidth]{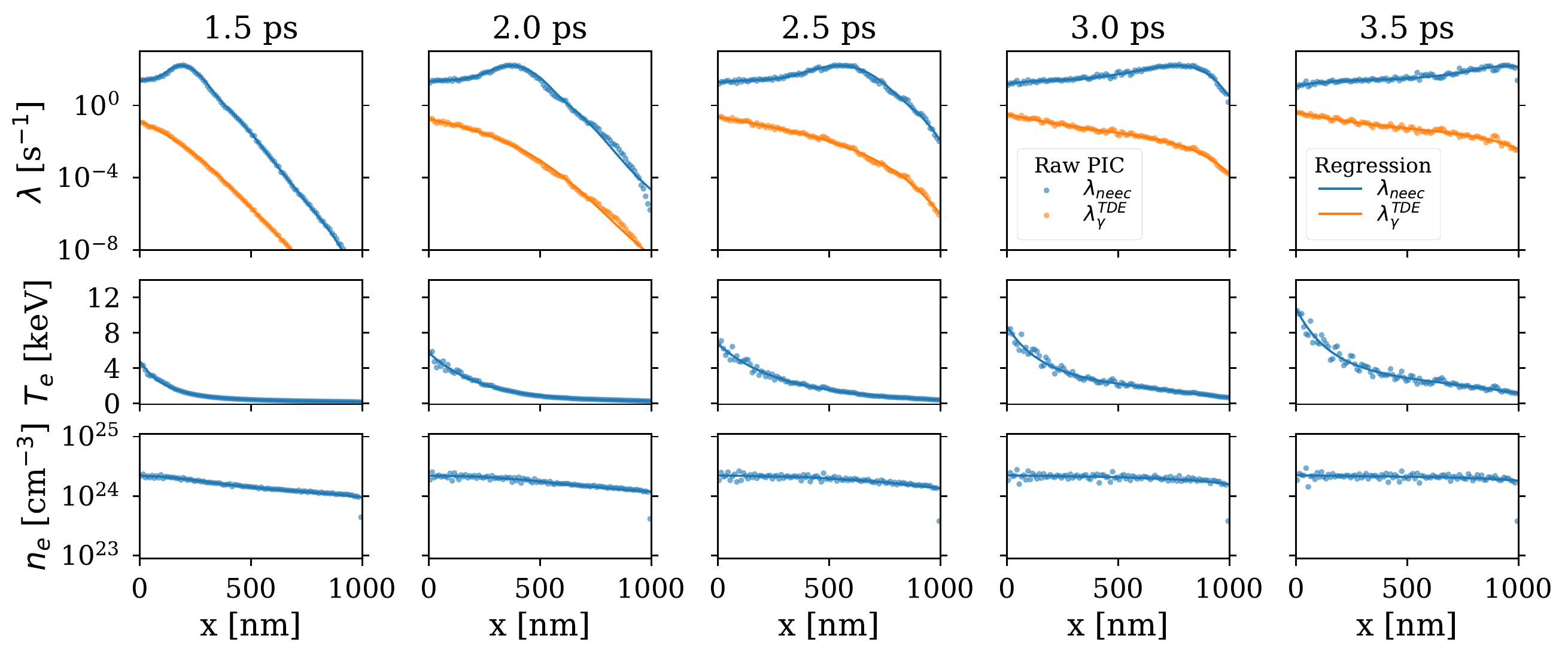}
    \caption{Nuclear excitation rate $\lambda$ (upper row), plasma temperature $T_{\text{e}}$ (middle row) and electron density $n_{\text{e}}$ (lower row) based on the PIC simulation as functions of target depth $x$ at time instants 1.5 ps, 2.0 ps, 2.5 ps, 3.0 ps amd 3.5 ps. The NEEC rate $\lambda_{\text{neec}}$ [blue (dark gray)] is shown in the upper-row graphs together with the photoexcitation rate $\lambda_{\gamma}^{\text{TDE}}$ [orange (light gray)]. The laser has peak intensity $I = 10^{18}$ W/cm$^2$ and wavelength $\lambda = 800$ nm. The raw data averaged over 10 nm intervals is presented together with a linear polynomial and a third order exponential fit for $n_{\text{e}}$ and $T_{\text{e}}$, respectively. The raw result of $\lambda_{\text{neec}}$ and regression curves for $\lambda_{\text{neec}}$ calculated with the fitted $n_{\text{e}}$ and  $T_{\text{e}}$ functions are shown in the upper row graphs.
}
    \label{fig:pic}
\end{figure*}

To include the effect of atomic ionization and recombination events, we averaged the raw data for electron temperature $T_\text{e}$ and ion density $n_{\text{i}}$ from the PIC simulation over 10 nm intervals, and used these values as input for the radiative-collisional model implemented in FLYCHK \cite{FLYCHK2005} to obtain charge state distributions and (corrected) electron densities.
The electron density and temperature values are shown in the lower and middle panels of Fig.~\ref{fig:pic} for a number of time instants between 1.5 and 3.5 ps as a function of the target penetration depth $x$ together with first order polynomial and third order exponential fits, respectively.
We note that due to ionization effects the real temperature is expected to be different from the one obtained in the PIC simulation. We use the latter only as first approximation.

\subsubsection{NEEC excitation}
For  the high-density region, we evaluate the NEEC rate as a function of target depth $x$ and time $t$ by inserting the PIC-simulation results for $T_{\text{e}}$ and the corrected $n_{\text{e}}$ values into Eqs.~\eqref{eq:neec-plasma.total} and \eqref{eq:neec-plasma.partial}. The plasma is assumed to be homogeneous only in the plane perpendicular to the $x$ direction over the region of $A_{\text{focal}}$.
We consider a laser pulse energy of 100 J, which leads for the pulse duration and laser intensity adopted in the PIC simulation to a focal spot area of approximatively $2\times 10^{-4}$ cm$^2$.
Results for $\lambda_{\text{neec}}$ and $\lambda_{\gamma}^{\text{TDE}}$ are presented as a function of $x$ for five time points between  $t=1.5$ and $t=3.5$ ps in the upper panel of Fig.~\ref{fig:pic}.
The NEEC rate is maximized at depths $x$ where optimal plasma conditions are prevailing. The peak propagates through the target and disappears at around 4 ps as the target heating leads afterwards to temperatures exceeding the optimal value for NEEC. A detailed analysis of data sampled from 1 to 4 ps in 100-fs steps shows that the integrated NEEC rate reaches its maximum at 3.1 ps and drops roughly to half its value at 4 ps. Due to the high electron density, $\lambda_{\text{neec}}$ is much larger than the photoexcitation rate over the entire target.

\begin{figure}
    \includegraphics[width=\linewidth]{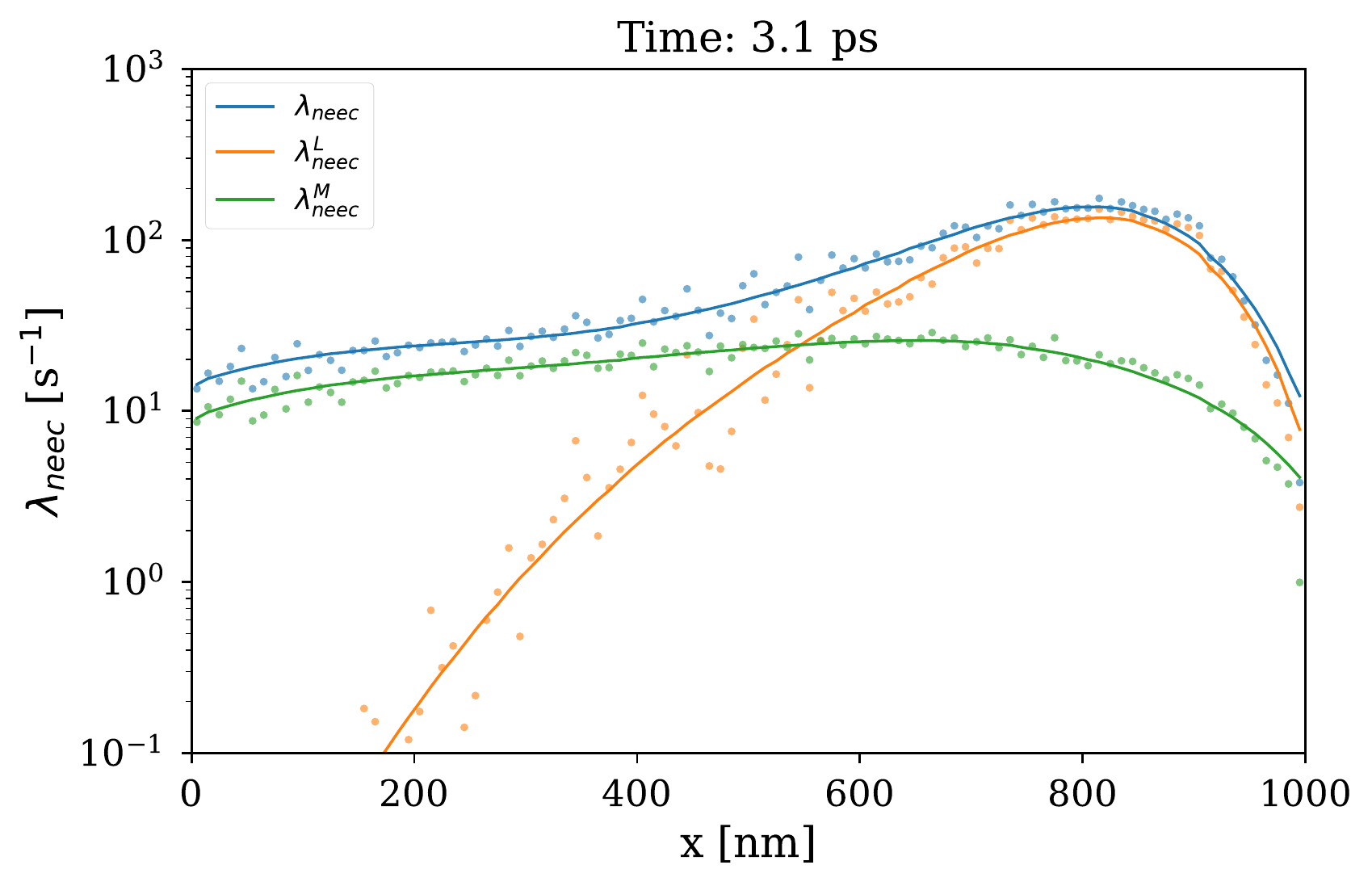}
    \caption{NEEC rate $\lambda_{\text{neec}}$ and regression curves for $\lambda_{\text{neec}}$ calculated with the fitted $n_{\text{e}}$ and $T_{\text{e}}$ functions are shown for the total rate [blue (dark gray)] as well as individual $L$-shell [orange (light gray)] and $M$-shell [green (medium gray)] contributions as a function of target depth $x$.}
    \label{fig:pic_shells}
\end{figure}

The total NEEC rate is shown together with its individual $L$- and $M$-shell contributions $\lambda_{\text{neec}}^{\text{L}}$ and $\lambda_{\text{neec}}^{\text{M}}$, respectively, in Fig.~\ref{fig:pic_shells} at the time instant of 3.1 ps where the integrated value reaches its maximum. The figure shows that the flat region mainly comes from the capture into the $M$ shell, which is available (almost) over the whole temperature-density landscape.
In contrast, the $L$-shell NEEC orbitals are only accessible in a very limited region in terms of plasma conditions, leading to the peak in $\lambda_{\text{neec}}$ at a target depth $x$ where optimal conditions for $L$-shell NEEC are prevailing.

Using the regression curves for $\lambda_{\text{neec}}$ calculated with the fitted $n_{\text{e}}$ and $T_{\text{e}}$ functions, we solve Eq.~\eqref{eq:N_exc} in a two-step procedure to obtain the total NEEC excitation number $N_{\text{exc}}$.
First, for each time instant $t$ the product of the NEEC rate and the isomer density is integrated with respect to $x$ over the entire target thickness $d_{\text{t}}$ and multiplied by the focal spot area $A_{\text{focal}}$ to account for the perpendicular directions.
Second, the outcomes of the spatial integration are interpolated as a function of time leading to $N_{\text{exc}}(t)$  which is defined as the derivative with respect to time of the number of excited isomers from initial time $t_0$ (here 1 ps) up to $t$,
\begin{equation}
    N_{\text{exc}}(t) = \int_{V_{\text{p}}}\! d^3\bold{r}\  n_{\text{iso}}(\bold{r},t) \, \lambda_{\text{neec}}(T_{\text{e}}, n_{\text{e}};\bold{r},t) \ .
\end{equation}
The interpolation for $N_{\text{exc}}(t)$  is then inserted in the time integral in Eq.~\eqref{eq:N_exc} which is solved numerically.

For $t>4$ ps, we extrapolate $N_{\text{exc}}(t)$  assuming an exponential functional behavior initially following the slope at 4 ps.
The time integration starting from 1 ps converges approximatively after 10 ps, leading to an excitation number of $1.8$ isomers per pulse via NEEC which is almost identical with the best value at low densities obtained with the PETAL parameters and again 6 orders of magnitude higher than the excitation in the XFEL scenario discussed in Refs.~\cite{GunstPRL2014, GunstPOP2015}.
Notable here is that in the high-density case a 100 J laser available at many facilities around the world is competitive with a kJ-laser facility.

\subsubsection{Modeling of the plasma expansion}
\begin{figure}
    \includegraphics[width=\linewidth]{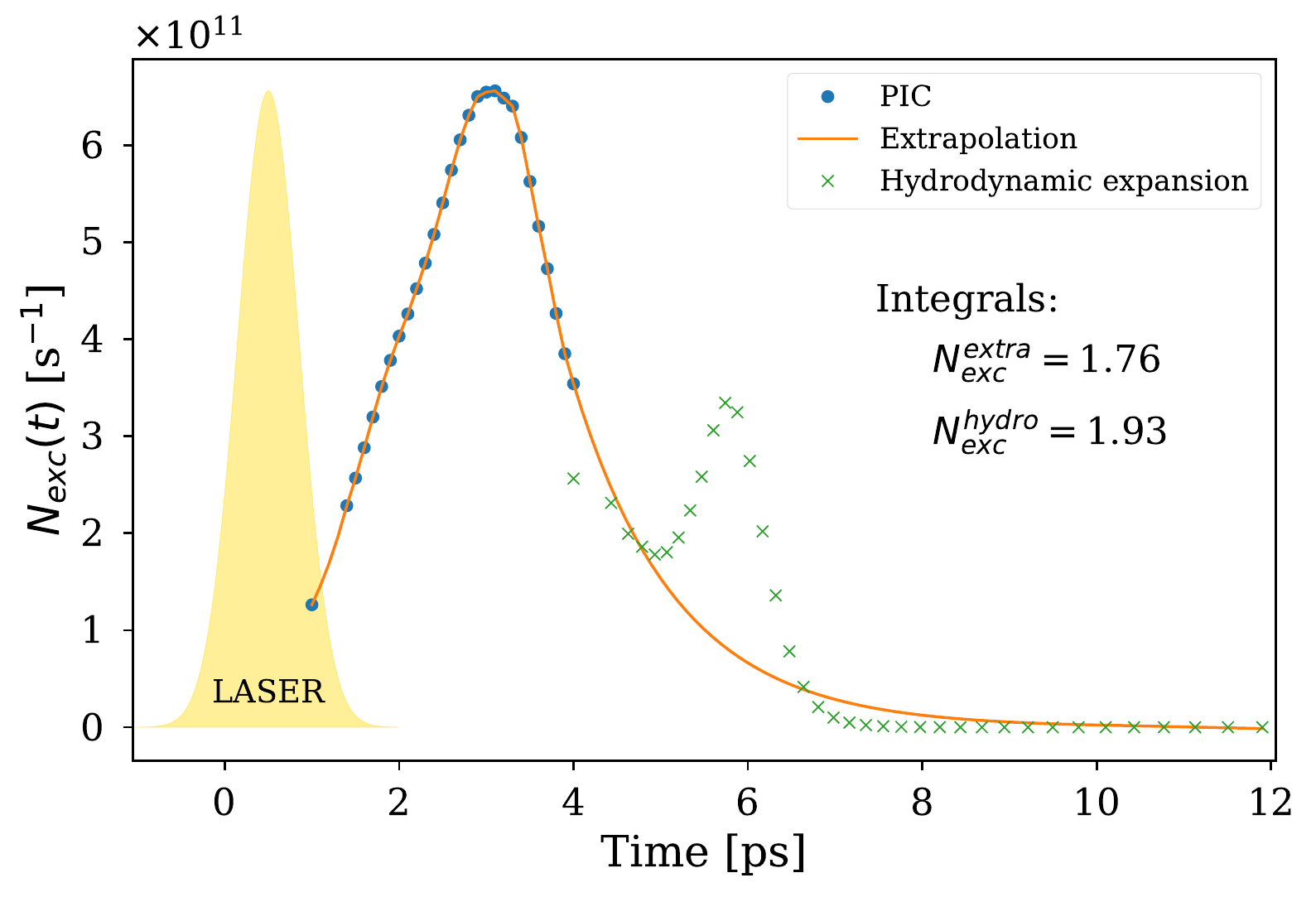}
    \caption{The time-dependent nuclear excitation number $N_{\text{exc}}(t)$ for PIC results (blue circles). For $t>4$ ps either an extrapolation (orange curve) or a hydrodynamic expansion (green crosses) are used to estimate $N_{\text{exc}}$. The laser pulse duration is schematically illustrated by the yellow area.}
    \label{fig:pic_expansion}
\end{figure}

The extrapolation method described above is equivalent with the assumption that the plasma heating continues after 4 ps. However, since no further energy is placed into the system, the plasma heating should reduce and finally turn into cooling during the plasma expansion. For a cross-check, we consider the plasma expansion to set in directly at 4 ps, and use a hydrodynamic model to estimate $N_{\text{exc}}$. We consider the average ion density and electron temperature at 4 ps as input for our hydrodynamic expansion model introduced in Section \ref{sec:hydro} assuming homogeneous plasma conditions over the plasma volume $V_{\text{p}}=A_{\text{focal}} d_{\text{t}}$. The results for the excitation number differential in time are shown in Fig.~\ref{fig:pic_expansion} for both the extrapolation as well as the expansion model.

During the cooling phase, $N_{\text{exc}}(t)$ reaches again a local maximum at the time where $T_{\text{e}} = T_{\text{max}}$ for the given density (see Fig.~\ref{fig:pic_expansion}).
However, since the density is also strongly decreasing during the expansion, the net effect for the excitation number is small.
Our calculations show that the extrapolation method as well as the hydrodynamic expansion deliver similar results for $N_{\text{exc}}$ with a deviation of 10\%. Moreover, the values from both methods are in good agreement (within a 20\% interval) with a simple lifetime estimate where according to Eq.~\eqref{eq:tau} the plasma lives for additional 2 ps with homogeneous plasma conditions given by the averaged values at $t = 4$ ps.

Note that the PIC simulation has been carried out in the direction with the smallest length scale of the plasma, such that our model underestimates the plasma lifetime and only gives a lower limit for the excitation number. Modeling the expansion in the perpendicular direction of the laser incidence with the length scale set by the focal radius, a 10 to 100-fold longer lifetime can be expected to boost $N_{\text{exc}}$.

\section{Conclusions}

Our results show that by a proper choice of target and optical laser parameters, the plasma conditions can be tailored to optimize  nuclear excitation via the NEEC process. For the case of
$^{93\mathrm{m}}$Mo, both  low-density and high-density plasmas promise observable depletion of the isomer. The induced excitation is expected to be six orders of magnitude higher than secondary NEEC in an XFEL-produced cold plasma and in turn a factor $10^{11}$ up to
$10^{12}$ higher than the direct photoexcitation with the XFEL. Allegedly the absolute number of depleted isomers remains small, mainly due to the fact that the number of isomers in the microscopic plasma volume is small and  only a $10^{-10}$  fraction of them gets depleted.

The excitation number of approximatively $2$ isomers per pulse from our conservative estimate together with laser repetition rates of up to tens of Hz for 100 J pulses  reach for the first time the threshold of one isomer depletion per second and should provide a detectable signal. The experimental signature of the nuclear excitation in the plasma would be a  gamma-ray photon of approx.~1~MeV  released in the decay cascade of the triggering level in $^{93}$Mo. An evaluation of the plasma black-body and bremsstrahlung radiation spectra at this photon energy shows that the signal-to-background ratio is very high.   An enhancement of the signal could be achieved by employing a combination of optical and x-ray lasers as envisaged for instance at HIBEF \cite{hibef} at the European XFEL \cite{europeanXFEL}.
X-rays-generated  inner shell holes could then provide the optimal capture state in the $L$-shell orbitals independently from the hot plasma conditions.

\bibliography{NEEC}

\end{document}